\documentclass[aps,prd,twocolumn,amsfonts,amssymb,amsmath,showpacs,letterpaper,superscriptaddress,nofootinbib,altaffilletter]{revtex4-1}

\usepackage[citecolor=blue,linkcolor=blue,colorlinks=true,urlcolor=blue]{hyperref}
\usepackage{color}
\usepackage{graphicx}
\usepackage{latexsym}
\usepackage{float}
\usepackage{amsmath}
\usepackage{amssymb}
\usepackage{multirow}
\usepackage{ifthen}
\usepackage{slashed}
\usepackage{lineno}
\usepackage{makecell} 
\usepackage[mathscr]{euscript} 
\usepackage{mathrsfs}

\newcommand{\hrss}{h_{\textrm{rss}}}

\newcommand{\cwb}{\textsc{cWB}}

\newcommand{\makevisible}[1]{\textcolor{red}{#1}}
\newcommand{\switch}[1]{%
  \ifthenelse{\equal{#1}{0}}{\renewcommand{\makevisible}[1]{}}{}}
\switch{1} 
\def\version$#1,v #2 #3${#2}

\usepackage{array}
\newcolumntype{C}[1]{>{\centering\let\newline\\\arraybackslash\hspace{0pt}}m{#1}}

\renewcommand{\today}{\number\day\space\ifcase\month\or
  January\or February\or March\or April\or May\or June\or
  July\or August\or September\or October\or November\or December\fi
  \space\number\year}

\def\be{\begin{equation}}
\def\ee{\end{equation}}
\def\bi{\begin{itemize}} 
\def\ei{\end{itemize}}
\def\ben{\begin{enumerate}}
\def\een{\end{enumerate}}
\def\hrssu{Hz${}^{-1/2}$}
\def\hrss{h_\mathrm{rss}}

\def\ligodoc{{LIGO-P17}{00177}{}}

\def\apj{Astrophys. J.}
\def\apjl{Astrophys. J. Lett.}
\def\apjs{Astrophys. J. Supp. Ser. }

\def\aap{Astron. Astrophys. }
\def\araa{Ann.\ Rev. Astron. Astroph. }

\def\mnras{Mon. Not. Roy. Astron. Soc. }

\def\prl{Phys. Rev. Lett.}
\def\prd{Phys. Rev. D.}

\begin{document}

\title{An Optically Targeted Search for Gravitational Waves emitted by Core-Collapse Supernovae during the First and Second Observing Runs of Advanced LIGO and Advanced Virgo}

  \let\mymaketitle\maketitle
  \let\myauthor\author
  \let\myaffiliation\affiliation
  \author{LIGO Scientific Collaboration}
  \author{Virgo Collaboration}
  \author{ASAS-SN Collaboration}
  \author{DLT40 Collaboration}
  \author{F.~Salemi}

\begin{abstract}

We present the results from a search for gravitational-wave transients associated
with core-collapse supernovae observed within a source distance of approximately 20\,Mpc
during the first and second observing runs of Advanced LIGO and Advanced Virgo.
No significant gravitational-wave candidate was detected.
We report the detection efficiencies as a function of the distance 
for waveforms derived from multidimensional numerical 
simulations and phenomenological extreme emission models.
For neutrino-driven explosions the distance at which we reach 50\% detection efficiency 
is approaching 5\,kpc, and for magnetorotationally-driven explosions is up to 54\,kpc.
However, waveforms for extreme emission models are detectable up to 28\,Mpc.
For the first time, the gravitational-wave data enabled us to exclude
part of the parameter spaces of two extreme emission models
with confidence up to 83\%, limited by coincident data coverage.
Besides, using \textit{ad hoc} harmonic signals windowed with Gaussian envelopes
we constrained the gravitational-wave energy emitted during
core-collapse at the levels of 
$4.27\times 10^{-4}\,M_\odot c^2$ and 
$1.28\times 10^{-1}\,M_\odot c^2$  
for emissions at 235\,Hz and 1304\,Hz respectively.
These constraints are two orders of magnitude more stringent than previously 
derived in the corresponding analysis using initial LIGO, initial Virgo 
and GEO~600 data.

\end{abstract}


\date[\relax]{Dated: \today }

\maketitle

\section{Introduction}
\label{sec:introduction}

The direct detection in September 2015 of a binary black hole merger~\cite{abbott:16a}
initiated the field of gravitational-wave astronomy.
During the first and second observing runs (O1 and O2) of Advanced LIGO and Advanced Virgo
several more mergers were reported~\cite{2018arXiv181112907T,2016PhRvX...6d1015A,2016PhRvL.116x1103A,2017PhRvL.118v1101A,2017PhRvL.119n1101A,2017ApJ...851L..35A} and 
in August 2017 a binary neutron star merger~\cite{GW170817} was observed 
in the gravitational-wave (GW) and electromagnetic spectra.
This event gave birth to multimessenger astronomy with gravitational 
waves~\cite{2017ApJ...848L..12A,LIGOScientific:2019gag,Monitor:2017mdv,2017Sci...358.1556C,Abbott2017hubble}.

Core-collapse supernovae (CCSNe) are another important target of multimessenger
astronomy with GWs, as all recorded supernovae were observed in the electromagnetic
spectrum and low energy neutrinos were observed from SN~1987A~\cite{hirata:87,bionta:87,1987ESOC...26..237A}.
GWs and neutrinos provide unique information about the 
dynamics of the collapse and the onset of the explosion,
as opposed to electromagnetic emission which is delayed and originates in 
regions thousands of kilometers away from the central engine. Their observation 
could provide hints to the shock revival mechanism~\cite{bhf:95,herant:95,couch:15,lentz:15,janka:12a,janka:17,mueller:16}.
The most promising opportunity for multimessenger GW astronomy with CCSNe
would be a Galactic CCSN, although the rate of such events is expected to be
just one or two per century~\cite{vandenbergh:91,cappellaro:93,tammann:94,li:11,diehl:06,maoz:10, 2013ApJ...778..164A}.

\begin{table*}[!bt]
\caption{Core-collapse supernovae selected as targets for the gravitational-wave search described in this paper.
The variables $t_{1}$ and $t_{2}$ are the start and end of the on-source windows (OSWs), $\Delta t$ is the duration of the OSWs, and 
OSW Method indicates how the OSW is calculated (see Sec.~\ref{sec:osw}).
The Run column indicates the LIGO and Virgo observing runs.
The Active Detectors column lists the interferometers taking data during the on-source window.
We include data from the LIGO Hanford (H1), LIGO Livingston (L1) and Virgo (V1) detectors.
The last column presents coincident duty factors.
}
\centering
\begin{tabular}{llcccccclrr}
\hline
\hline
Supernova & Type &\multicolumn{1}{c}{Host}&\multicolumn{1}{c}{Distance}
& \multicolumn{1}{c}{$t_{1}$} & \multicolumn{1}{c}{$t_{2}$}
& \multicolumn{1}{c}{$\Delta t$} & \multicolumn{1}{c}{OSW} & \multicolumn{1}{c}{Run} & \multicolumn{1}{c}{Active}&\multicolumn{1}{c}{Coincident}\\
& &\multicolumn{1}{c}{Galaxy} &\multicolumn{1}{c}{[Mpc]} &\multicolumn{1}{c}{[UTC]} 
&\multicolumn{1}{c}{[UTC]} & \multicolumn{1}{c}{[days]} & \multicolumn{1}{c}{Method} &\multicolumn{1}{c}{} & \multicolumn{1}{c}{Detectors}&\multicolumn{1}{c}{Coverage}\\
\hline
SN 2015as & IIb & UGC 5460   &$19.2$ & 2015 Nov 14.77 & 2015 Nov 16.23 & 1.47 & Early & O1 & H1,L1 & 34.2\%\\
SN 2016B  & IIP & PGC 037392 &$19.1$ & 2015 Dec 23.51 & 2015 Dec 27.55 & 4.03 & Early & O1 & H1,L1 & 34.3\%\\
SN 2016X  & IIP & UGC 08041  &$15.2$ & 2016 Jan 17.72 & 2016 Jan 20.56 & 2.86 & Early & O1 & H1,L1 & 14.4\%\\
SN 2017eaw & IIP & NGC 6946  &$6.72$  & 2017 Apr 26.56 & 2017 Apr 27.96 & 1.39 & EPM & O2 & H1,L1 & 48.8\%\\
\multirow{2}{*}{SN 2017gax} & \multirow{2}{*}{Ib/c} & \multirow{2}{*}{NGC 1672} & \multirow{2}{*}{$19.7$} 
  & \multirow{2}{*}{2017 Aug 14.28} & \multirow{2}{*}{2017 Aug 16.15} & \multirow{2}{*}{1.66} & \multirow{2}{*}{Early} & \multirow{2}{*}{O2} & \multirow{2}{*}{H1,L1,V1} & 61.5\% (H1L1)\\
          &     &        &   &          &                   &       &     &   &     & 60.8\% (H1L1V1)\\
\hline
\hline
\end{tabular} 
\label{tab:sne}
\end{table*}

\begin{figure}[b]
\centering
\includegraphics[width=1.00\columnwidth]{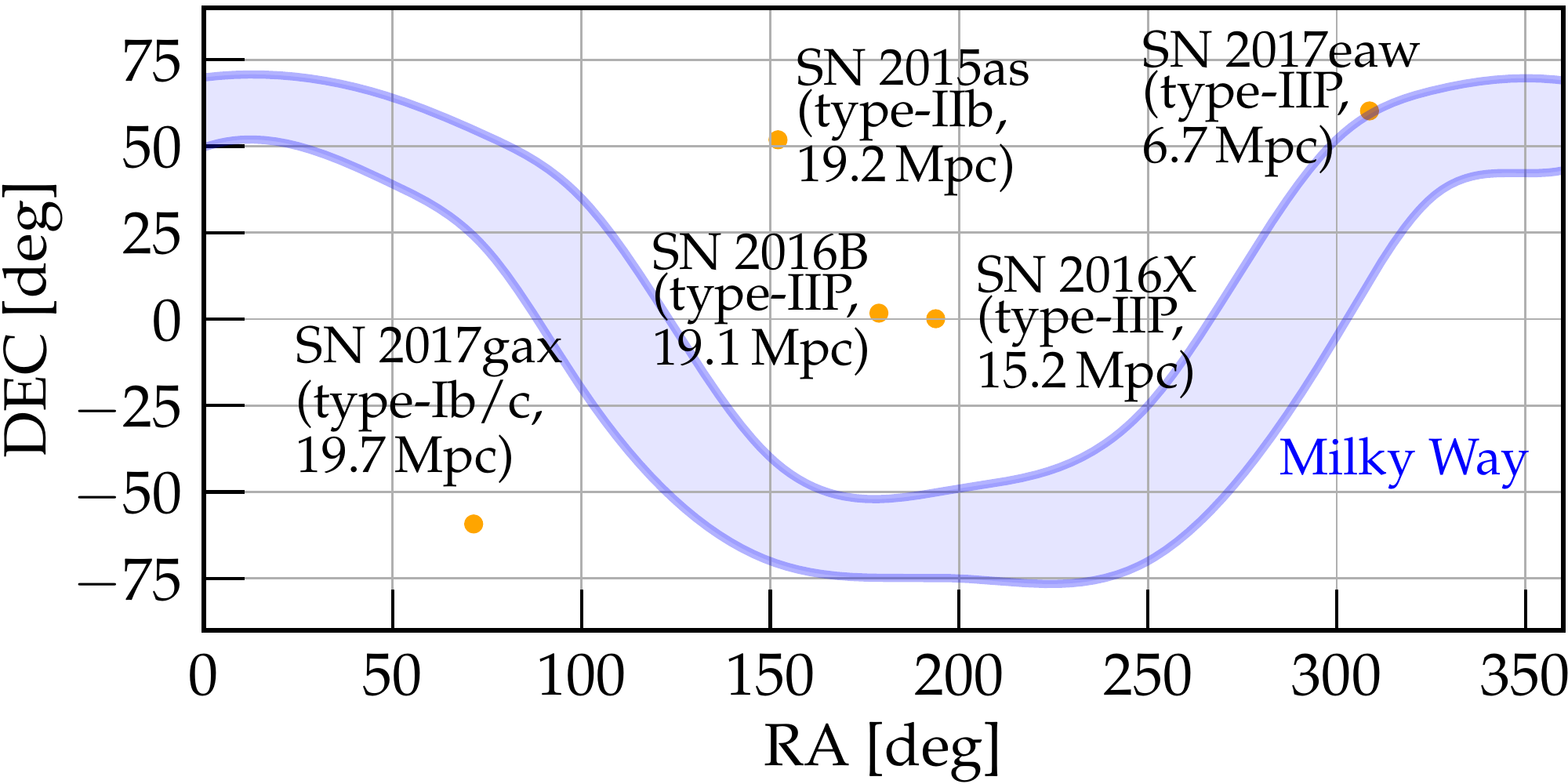}
\caption{
Sky locations of core-collapse CCSNe analyzed in this search. 
All were recorded within 20\,Mpc during the O1 and O2 observing runs.
}
\label{fig:sky_loc}
\end{figure}

In contrast to all-sky, all-time unmodelled GW transient 
searches~\cite{abadie:12s6burst,ligo_burst_s5y1:09,abbott17,PhysRevD.100.024017}, targeted searches
for CCSNe impose the sky location, the source distance, and a time window for the arrival time of the GW signal.
In the previous CCSN targeted search with first-generation GW detector data~\cite{snsearch}
we developed the methodology, derived distance ranges
for various GW emission processes, provided null model exclusion statements,
and established GW energy constraints.

This paper describes a targeted search focusing on CCSNe recorded by astronomical 
observations at distances up to approximately 20\,Mpc during O1 and O2.
We selected five CCSNe, four of which are type-II supernovae 
(SN~2015as, SN~2016B, SN~2016X, SN~2017eaw) and one is
type-Ib/c (SN~2017gax).
We have not found any evidence for a GW signal associated with them.
Similarly to~\cite{snsearch}, we obtain distance ranges
for a selection of waveforms which were computed from
numerical simulations and that are representative of different
emission mechanisms and progenitors.
We also use phenomenological waveforms
representing possible but extreme emission models and
we derive standard candle model exclusion statements for them.
Finally, we adopt ad hoc sine-Gaussian waveforms to
simulate GW emission in specific time-frequency regions
allowing us to derive upper limits on the emitted GWs from a specific CCSN.

This paper is organized as follows:
In Sec.~\ref{sec:targeted_sne} we list the CCSNe that we study in this search. We also describe
methods for calculating the time period when we expect the moment of collapse.
In Sec.~\ref{sec:gw_network} we describe the data used in the search.
Sec.~\ref{sec:methodology} describes the methodology, the pipeline, simulated GW signals,
and systematic uncertainties.
The results in Sec.~\ref{sec:results} include distance reaches for several models of emission,
GW energy constraints, and model exclusion statements.
We draw conclusions in Sec.~\ref{sec:summaries}.

\section{Targeted Core-Collapse Supernovae} 
\label{sec:targeted_sne}

From all core-collapse supernovae recorded during the O1 and O2 periods,
we have selected those that contribute to model exclusion statements
and meet the following criteria:
(i) the distance is less than approximately 20\,Mpc, 
(ii) the period where we expect to find the GW transient, the on-source window, (see Sec.~\ref{sec:osw}) is sufficiently well identified (order of days maximum), and  
(iii) there is sufficient GW detector data within the supernova on-source 
window to allow us to accumulate at least a few years of background data (see Sec.~\ref{sec:bkg}).

During O1 and O2, astronomers 
found and followed-up numerous CCSNe in the nearby universe.  
Based on the information from 
Astronomical Telegrams~\cite{Rutledge1998} and supernova catalogs
(ASAS-SN~\cite{Shappee2014, 10.1093/mnras/stz073, 10.1093/mnras/stx1544, 10.1093/mnras/stx057}, DLT40~\cite{2017ATel10638....1T}, 
Gaia~\cite{2016A&A...595A...1G,2016A&A...595A...2G}, ASRAS~\cite{www:asras}, 
TNS~\cite{2018ApJ...853...62T}, OSC~\cite{2017ApJ...835...64G}, CBAT~\cite{www:cbat}), we found 9 supernovae of interest.

Only five CCSNe meet the selection criteria and are used for the astrophysical statements in this paper. They are:
SN~2015as, SN~2016B, SN~2016X, SN~2017eaw, and SN~2017gax.
They are reported in Table~\ref{tab:sne} and
Figure~\ref{fig:sky_loc} presents their sky locations.
The majority of these are type-II supernovae originating from red supergiant progenitor stars and the host galaxy was identified for each. 
The distance to each CCSNe is determined using the estimated 
distance to its host galaxy.  
The on-source window calculation methods are described Sec.~\ref{sec:osw}.

\begin{figure*}[t] 
\centering 
\includegraphics[width=0.70\textwidth]{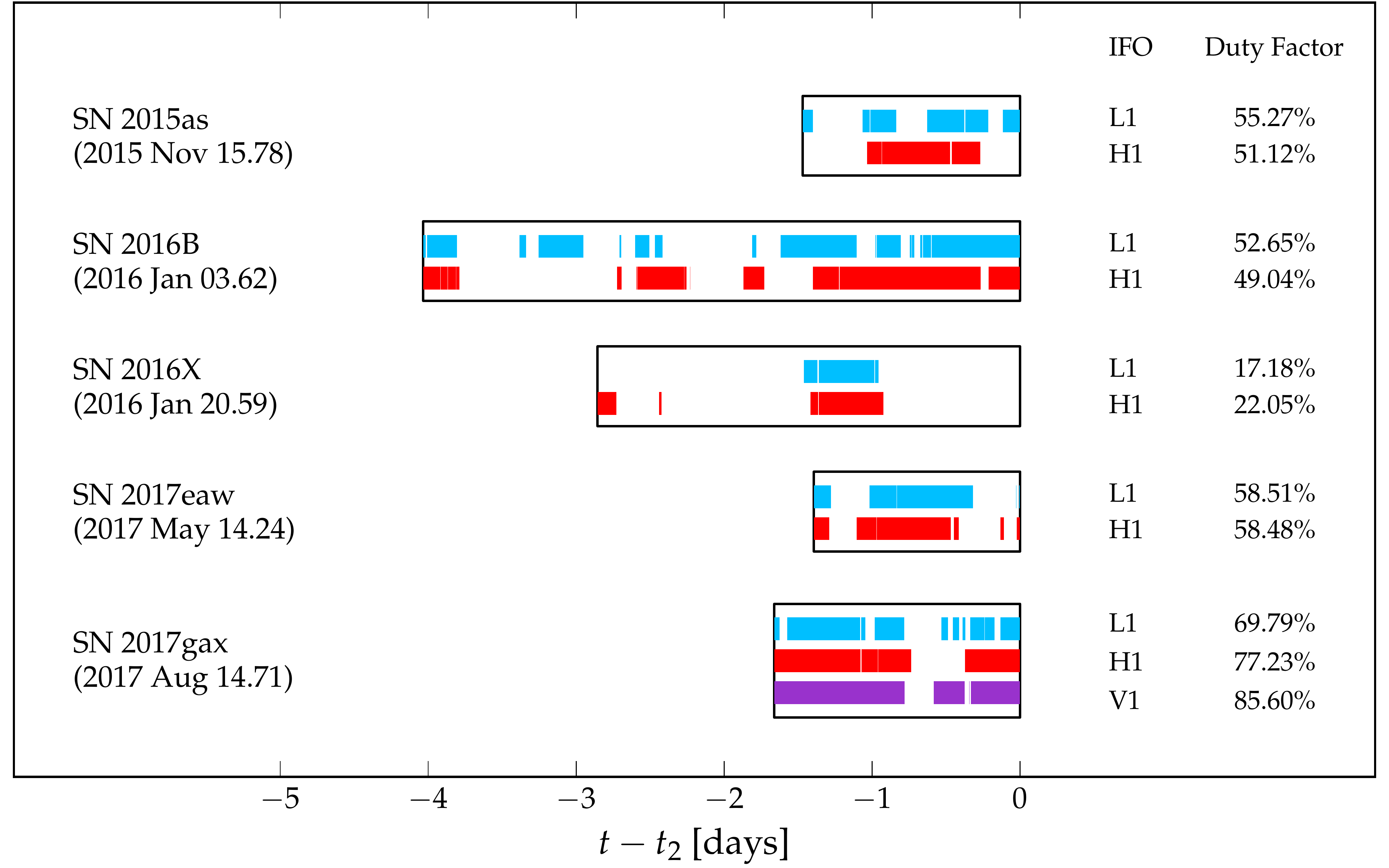} 
\caption{Visual representation of the on-source windows (see Sec.~\ref{sec:osw}), 
the data coverage for each detector, and the detector
duty factors (percentage of available data inside on-source window). 
Dates in brackets are CCSN discovery times in UTC and 
$t_2$ is the end time of the on-source window for each CCSN.
The plotted interferometers (IFO) are LIGO Hanford (H1), LIGO Livingston (L1) and Virgo (V1).
} 
\label{fig:coverage} 
\end{figure*}

\textit{SN~2015as}, a type-IIb supernova, was discovered 
on 2015 November 15.78 UTC~\cite{2015ATel.8291....1T} during O1. 
The host galaxy is UGC~5460 at a distance of 19.2\,Mpc~\cite{1802.06731}. 
Although the spectrum transitions to a type-Ib supernova around 75 days after explosion, 
the spectrum evolution closely relates to that of SN~2008ax, suggesting type-IIb~\cite{1802.06731}. 
The progenitor star is either a main sequence $15\,M_\odot$ star 
or $20\,M_\odot$ Wolf-Rayet star~\cite{2018BSRSL..87..351G}. 
CCSN ejecta is estimated to be $1.1-2.2\,M_\odot$.

\textit{SN~2016B} (ASASSN-16ab), a type-IIP supernova, was discovered by ASAS-SN on 2016 January 03.62 UTC~\cite{2016ATel.8502....1S} 
during O1. The host galaxy is PGC~037392 at a distance of 18.6\,Mpc~\cite{2016ATel.8502....1S}.
The progenitor star is estimated to be a red supergiant~\cite{2018MNRAS.479.3232G}.

\textit{SN~2016X} (ASASSN-16at), a type-IIP supernova, was discovered by ASAS-SN on 2016 January 20.59 UTC~\cite{2016ATel.8566....1B}.
It exploded in the spiral galaxy UGC~08041 at a distance of 15.2\,Mpc~\cite{Huang2018}. 
Optical observations in~\cite{Huang2018} indicate that the 
progenitor star is a massive red supergiant with an initial mass larger than $19-20\,M_\odot$ 
and a radius larger than $930\,\pm\,70\,R_\odot$.

\textit{SN~2017eaw} (Gaia17bmy), a type-IIP supernova, was discovered by Gaia
on 2017 May 14.24 UTC~\cite{2017ATel10372....1D}. The CCSN exploded 
in galaxy NGC~6946, the estimated distance to be $6.72\pm0.15$\,Mpc away~\cite{2014AstL...40..537T}.
This is the closest CCSN considered in the search. 
The analyses in~\cite{Kilpatrick2018, Tsvetkov2018, tinyanont:19} provide indication that the progenitor was a red supergiant
with an estimated initial mass of $13\,M_\odot$ and radius of $4000\,R_\odot$.  
 
\textit{SN~2017gax} (DLT17ch), a type-Ib/c supernova, was discovered by the DLT40
on 2017 August 14.71 UTC~\cite{2017ATel10638....1T}.  
This CCSN was found in NGC~1672, 19.7\,Mpc away~\cite{www:sn2017gax}. 
Unfortunately, little is known about the progenitor star.

Any CCSN, where the detection efficiencies for the extreme
emission models are non-zero, and with sufficient on-source window coverage,
helps the model exclusion probabilities (see Sec.~\ref{sec:mep}).
In this regard we also considered CCSNe at distances greater than 20\,Mpc.
Four other such CCSNe have been recorded during the O1 and O2 periods:
not enough GW data was available for \textit{SN~2016C} (type-IIP, 20.1\,Mpc~\cite{www:sn2016C, 2016ATel.8514....1S}) 
and \textit{SN~2017ein} (type-Ic, 11.2\,Mpc~\cite{www:SN2017ein, 2019ApJ...871..176X}),
and no on-source window could be sufficiently constrained for \textit{SN~2017aym} (Gaia17aks) (type-IIP, 26.4\,Mpc~\cite{www:SN2017aym, 2017TNSTR.214....1D})
and \textit{SN~2017bzb} (type-II, 13.9\,Mpc~\cite{www:SN2017bzb, 2017TNSTR.279....1P}). 
All the other CCSN candidates occured outside the O1 and O2 periods or were located further than 20\,Mpc.

\begin{table*}[t] 
\caption{Overview of GW interferometers for the O1 and O2 observing runs from which we draw 
data for our search.  
The O1 observing run lasted four months and was followed by a half year maintenance period. 
The observing runs were preceeded by engineering runs which we do not report here.
O2 lasted around nine months, however the run was interupted twice, between 
2016.12.22--2017.01.04 and 2017.05.08--2017.06.26.
The numbers in the table were calculated after periods of poor data quality were removed.
} 
\begin{tabular}{l@{\hspace{0.3em}}c@{\hspace{1em}}c@{\hspace{1em}}c@{\hspace{1em}}c} 
\hline 
\hline 
Run&Detectors&Run Period&Duty Factors& Coin.\ Duty Factor\\ 
\hline 
O1&H1,L1  &2015.09.12--2016.01.19& 49.5\% (H1), 42.4\% (L1)& 31.4\% (H1L1)\\ 
O2&H1,L1   &2016.11.30--2017.08.25& 65.4\% (H1), 63.6\% (L1)& 49.0\% (H1L1)\\ 
O2&H1,L1,V1&2017.08.01--2017.08.25& 77.7\% (H1), 79.2\% (L1), 85.1\% (V1)& 62.0\% (H1L1V1)\\ 
\hline 
\hline 
\end{tabular} 
\label{tab:runs} 
\end{table*} 
 
\subsection{On-source window calculation} 
\label{sec:osw} 

The collapse of a star's iron core forms a proto-neutron 
star and initiates a hydrodynamical shock wave propagating outward.​
Depending on the size of the progenitor star, 
the ensuing shock propagates for a period of seconds to days~\cite{waxman:17}.
When it reaches the surface, i.e shock breakout, a CCSN emits observable light.
Because of weather conditions, limited sky coverage, and many other limitations,
astronomical surveys typically record CCSNe hours to months after light first reaches Earth.
The ability to extrapolate backwards in time to the moment of core-collapse,
depends primarily on
how quickly a CCSN is detected, its last non-detection, and the properties of its progenitor star.

The on-source window (OSW) is defined as the time interval $[t_1,t_2]$,
where $t_1$ and $t_2$ are the beginning and end times respectively.
An upper bound to this interval is $t_\mathrm{disc}$, the time at which the 
CCSN was first observed electromagnetically. We define $t_\mathrm{Null}$ as the time 
of the last observation of a host galaxy without a supernova present.
To estimate the OSW we consider two methods. 
The choice between the \textit{early observation method (Early)} 
and the \textit{expanding photosphere method (EPM)} is based on the 
quality of the multi-band photometry, the determination of the host galaxy and the type of CCSN.
We apply the early observation method when $t_\mathrm{disc}-t_\mathrm{Null}$ is of order a few days, the supernova type is known, and
the progenitor star is inferred~\cite{GalYam2014,Smartt2002,VanDyk2017,eldridge:06}.
In all other cases, we consider the EPM.

In the early observation method, $t_2$ is the time when the CCSN is discovered, i.e. $t_2=t_\mathrm{disc}$.
To determine $t_1$\,, we need to take into account $t_\mathrm{Null}$\,, and the shock propagation travel time
between the moment of explosion and shock breakout, $\Delta t_\mathrm{SB}$\,. 
We get that $t_1 = t_\mathrm{Null} - \Delta t_\mathrm{SB}$.
$\Delta t_\mathrm{SB}$ depends mainly on the type of the progenitor star.
Wolf-Rayet stars are stripped of helium and hydrogen and they lead to type-Ib/c supernovae.
Their radii are on the order of a few $R_\odot$ with typical shock breakout times ranging
from a few seconds up to a minute~\cite{Li2007}.
Red supergiant stars have radii of 500-1000\,$R_\odot$~\cite{smartt:09b}
and typical $\Delta t_\mathrm{SB}$ ranges from more than ten hours up to a few days~\cite{Davies2017}.
We calculated the OSW with the early observation method for
four CCSNe: SN~2015as, SN~2016B, SN~2016X and SN~2017gax.
For each of them we identified $t_\mathrm{Null}$ and $t_\mathrm{disc}$ based on the astronomical surveys.
We calculated $\Delta t_\mathrm{SB}$ from information about their progenitor stars.
To account for uncertainties in the progenitor star information, we enlarge 
the OSW by a number of hours (from 15\,h up to 24\,h~\cite{2008Sci...321..223S,2017A&A...605A..83D}).

The expanding photosphere method is used in astronomy primarily to estimate distances
distances to CCSNe, but we employ it to estimate the time of core-collapse~\cite{gilldcc:15}.
We briefly describe the method, but a detailed explanation can be found 
in~\cite{Waxman2017,2007AIPC..937..394V,2014ApJ...782...98B,2011AAS...21733721E,2012AIPC.1429..204L,2015ApJ...814...63M}.
When the shock breaks through a star's surface, it heats up the outer layers and pushes them outwards.
The expanding photosphere grows with time and its speed can be measured using
Doppler shifts in its spectrum.
As a consequence, we can extrapolate backward in time from the moment when explosion was caught in the optical bands in order to estimate $t_1$ and $t_2$.
Since the interval between $t_\mathrm{disc} - t_\mathrm{Null}$ for SN~2017eaw was greater than a week
and because SN~2017eaw follow-up observations allow it~\cite{Kilpatrick2018, Tsvetkov2018},
we used the EPM to calculate the OSW.

\begin{figure}[hbt] 
\centering 
\includegraphics[width=1.0\columnwidth]{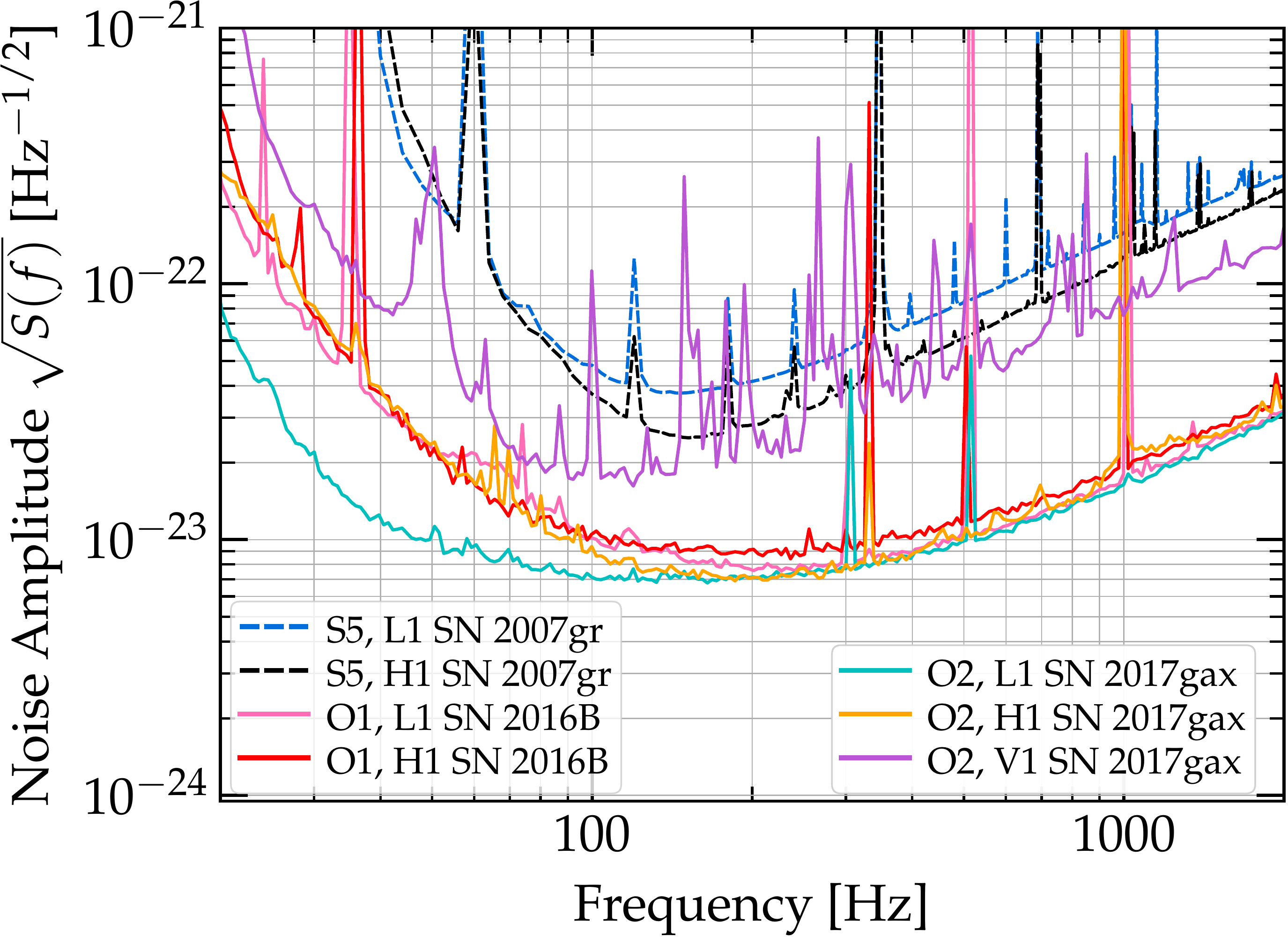} 
\caption{Noise amplitude spectral densities of the GW interferometers.  
For SN~2016B and SN~2017gax we chose ten random periods inside the
corresponding on-source windows. Each period was 10~minutes long.
We calculated the noise spectra for each and then took an average. 
Amplitude spectra for SN~2007gr are reproduced from Ref.~\cite{snsearch}.
  } 
\label{fig:noise} 
\end{figure}

\section{Detector Networks and coverage}
\label{sec:gw_network}

Data from O1 and O2 were used for this search.
This includes data from the Advanced LIGO detectors in Hanford (H1) and Livingston (L1) and the Advanced Virgo (V1) detector.
O1 started on September 12, 2015 and ended on January 19, 2016, 
while O2 spanned the period between November 30, 2016 and August 25, 2017~\cite{2018arXiv181112907T}.
The L1 and H1 detectors were observing jointly during O1 and O2,
and Virgo joined the LIGO detectors on August 1, 2017~\cite{2018arXiv181112907T}.

The data are affected by instrumental and environmental sources of noise that 
prevent some of the data from being analyzed~\cite{abbott:16a-detchar}. 
Periods of poor data quality are excluded using the information from 
many probes monitoring the environment of the detectors and probes 
controlling the different optical elements~\cite{Abbott2016min,Nuttall2018}.
For all CCSNe, we applied the same criteria for excluding times of poor data quality
that were used in the all-sky short-duration unmodelled transient GW searches~\cite{abbott17,PhysRevD.100.024017}.

Table~\ref{tab:runs} specifies the data taking periods along 
with the availability of each detector, which is referred to as the \textit{duty factor}.
The duty factor for each detector in O2 was higher than in O1 and was growing towards the end of the period.
Figure~\ref{fig:coverage} shows the OSWs for each CCSN
together with the periods when detector data was available.
The analysis is performed on data that is coincident between two or more detectors.
OSW for SN~2016X extends past the end of O1.

Figure~\ref{fig:noise} shows the noise spectral density
for all available detectors at the times of SN~2016B and SN~2017gax.
For comparison, we plot the sensitivities of the two detectors for SN~2007gr~\cite{snsearch} observed during LIGO Science Run~5~(S5).
When comparing the data for SN~2007gr to the O1-O2 sensitivity, the detectors are 3 to 5 times more sensitive in the most sensitive detector band
between 100 and 300\,Hz~\cite{abbott17}
and around 10 times more sensitive around 1\,kHz.
Moreover, the low frequency part of the spectrum (below 100\,Hz) improved during O2,
especially in the L1 detector.

SN~2017gax happened in August 2017 when the LIGO and Virgo detectors 
were acquiring data. We have considered not only the H1L1 coincident time 
but also the H1L1V1 coincident time. Because of the sensitivity difference 
between Virgo and the two LIGO detectors, we found that the H1L1V1 network 
is less sensitive. We thus report results using the H1L1 network.

\section{Methodology}
\label{sec:methodology}

\begin{table*}
\caption{
  Waveforms from detailed multidimensional CCSN
  simulations described in the text. 
  For each waveform, we give the
  emission type, reference, waveform identifier,
  angle-averaged root-sum-squared strain $h_\mathrm{rss}$, 
  the frequency $f_\mathrm{peak}$ at which the GW energy spectrum peaks, the
  emitted GW energy $E_\mathrm{GW}$, and available polarizations.
  See Refs.~\cite{gossan:16,szczepanczykdcc:15} for details.
  }
\begin{tabular}{clcrcc}
\hline
\hline
\multicolumn{1}{c}{Waveform}
&\multicolumn{1}{l}{Waveform}
&\multicolumn{1}{c}{$h_\mathrm{rss}$}
&\multicolumn{1}{c}{$f_\mathrm{peak}$}
&\multicolumn{1}{c}{$E_\mathrm{GW}$}
&\multicolumn{1}{c}{Polarizations} \\
Family 
& Identifier
&\multicolumn{1}{c}{$[10^{-22}\sqrt{\mathrm{s}}\,@ 10\,\mathrm{kpc}]$}
&\multicolumn{1}{c}{$[\mathrm{Hz}]$}
&\multicolumn{1}{c}{$[10^{-9} M_{\odot} c^{2}]$}\\
\hline
\hline
\multirow{3}{*}{\makecell{M\"uller \cite{mueller:e12}\\ 3D Convection and SASI}}
&mul1-L15-3 &1.655 &150 &\phantom{0}$3.741\times 10^{-2}$ &$+$, $\times$\\
&mul2-N20-2 &3.852 &176 &\phantom{0}$4.370\times 10^{-2}$ &$+$, $\times$\\
&mul3-W15-4 &1.093 &204 &\phantom{0}$3.247\times 10^{-2}$ &$+$, $\times$\\
\hline
Ott~\cite{ott:13a} 3D Convection and SASI
&ott1-s27fheat1p05 &0.238 &$1019$ &\phantom{0}$7.342\times 10^{-1}$ &$+$, $\times$ \\
\hline
\multirow{4}{*}{\makecell{Yakunin \cite{yakunin:15}\\2D Convection and SASI}}
&yak1-B12-WH07 &3.092 &760 &3.411 &$+$\\
&yak2-B15-WH07 &14.16 &932 &7.966 &$+$\\
&yak3-B20-WH07 &3.244 &638 &4.185 &$+$\\
&yak4-B25-WH07 &18.05 &1030&14.21 &$+$\\
\hline
\multirow{3}{*}{\makecell{Scheidegger~\cite{scheidegger:10b}\\ Rotating Core-Collapse}}
&sch1-R1E1CA$_L$ &0.129 &1155 &$1.509\times 10^{-1}$ &$+$, $\times$\\
&sch2-R3E1AC$_L$ &5.144 &466  &$2.249\times 10^{2}$ &$+$, $\times$\\
&sch3-R4E1FC$_L$ &5.796 &698  &$4.023\times 10^{2}$ &$+$, $\times$\\
\hline
\multirow{3}{*}{\makecell{Dimmelmeier~\cite{dimmelmeier:08}\\ Rotating Core-Collapse}}
&dim1-s15A2O05ls &1.052 &770 &\phantom{0}\phantom{0}7.685 &$+$\\
&dim2-s15A2O09ls &1.803 &754 &\phantom{0}27.880 &$+$\\
&dim3-s15A3O15ls &2.690 &237 &\phantom{0}\phantom{0}1.380 &$+$\\
\hline
\hline
\end{tabular}
\label{tab:models1}
\end{table*}

In this paper we search for GW signals in a large frequency band, 16-2048\,Hz, 
without specific assumptions about the signal morphology.
This frequency band allows us to cover most of the main emission processes inside a CCSN.
We employ coherent WaveBurst (\cwb)~\cite{cwb} as the search algorithm, 
which we describe in the following section.

\subsection{Coherent Waveburst}
\label{sec:cwb}

Coherent WaveBurst is an excess power pipeline that is based on the constrained maximum likelihood ratio method~\cite{cwb}.
For each event, the pipeline calculates correlation coefficients 
$cc=E_c / (E_c + E_n)$ which measures 
the degree of similarity of the waveforms between the detectors.
$E_c$ is the normalized coherent energy obtained by cross-correlating
the reconstructed waveforms in each detector and $E_n$ is the 
normalized per detector residual noise energy after the reconstructed waveform is
subtracted from the data.
For a real GW, $cc\approx1$, and we accept events
that have $cc>0.8$.
Each event is ranked according to a coherent network signal-to-noise ratio,
$\rho \propto \sqrt{E_c}$.
A more detailed explanation of these statistics is given
in Refs.~\cite{Abbott2016min,cwb}.

The events are divided into two mutually exclusive classes
based on their morphologies, similarly to~\cite{abbott17}.
Class~\textit{C1} contains transients of a few cycles.
This class is primarily polluted by \textit{blip} glitches
which are very short duration transients, $O(10)$\,ms, with large
bandwidth, $O(100)$\,Hz~\cite{abbott:16a-detchar,Cabero:19}.
These noise transients are currently of unknown origin.
To separate blip glitches from the bulk of the trigger population in class~\textit{C2} we use the selection 
criteria described in~\cite{Abbott2016min}.

\subsection{Background estimation}
\label{sec:bkg}

As mentioned earlier, each GW detector is constantly monitored 
with various sensors that allow us to exclude poor data quality periods from the analysis. 
However, it is not possible to remove all sources of noise.
To estimate how often the pipeline produces events that are falsely identified
as GWs, we perform a background analysis 
where \cwb~artificially shifts the data in one detector with respect to the other.
The typical time shift is a multiple of one second, which is much longer than the 
GW travel time between different detectors (e.g., 10\,ms between H1 and L1). 
This allows us to estimate the false alarm rate (FAR) of the background events.
We use a few years of background data for each CCSN.

GW events obtained when no shift is applied to the data may contain genuine GW signals.
The events from the search classes are combined and ranked with their FAR.
We assume that the event with the smallest FAR between the two search
classes is a potential GW candidate and we refer to it as
the \textit{loudest event}. The FAR is calculated from the noise
transient distribution of the class to which the loudest event
belongs. Since the classes are independent, we apply
a trial factor 2 to the FAR (see also~\cite{abbott17, abbott:16a}) of the
loudest event. The significance of an event with given FAR is assessed
by calculating its False Alarm Probability (FAP), which is the probability of 
obtaining one or more noise events that are less than or equally ranked:
\begin{equation}
  \label{eqn:fap}
  \mathrm{FAP} = 1-\exp{(-T_\mathrm{coinc} \times \mathrm{FAR})} \,
\end{equation}
where $T_\mathrm{coinc}$ is the coincident data duration of the appropriate OSW.

\subsection{Search sensitivity}
\label{sec:sim}

We determine how sensitive the pipeline is to particular waveform families.
\cwb~adds (\textit{injects}) supernova waveforms to the detector data inside the 
OSW with the right time delay in each detector such that the GW signal comes from the accurately known CCSN sky location.
The fraction of the injected signals that can be detected and pass the selection criteria 
is the \textit{detection efficiency}. The injection procedure is repeated with waveform amplitudes corresponding
to different source distances.
We select any event that passes the selection criteria of the search 
and whose rank is smaller than the loudest event FAR.

We consider two sets of multidimensional 
supernova explosion models, extreme emission models, and \textit{ad hoc} waveforms as listed in Tables~\ref{tab:models1} and~\ref{tab:models2}.
For all of the waveforms we provide the peak frequency, number of polarizations, and other quantities.
For the waveforms from multidimensional CCSN simulations and extreme
emission models, we provide the GW energy, $E_\mathrm{GW}$, emitted during the explosion
and the source angle-averaged root-sum-squared GW strain,
\begin{equation}
\label{eq:hrss}
\hrss=\sqrt{\int \left\langle h^2_+(t) + h^2_\times(t) \right\rangle_\Omega \mathrm{d}t}\, .
\end{equation}

Our efficiency estimates are subject to a number of uncertainties.
The most important of these are calibration uncertainties in the strain data
recorded at each detector, and Poisson uncertainties due to the use of a finite number
of injections (Monte Carlo uncertainties). We use the same
methodology explained in~\cite{snsearch}, to account for each of these uncertainties.
For detection efficiencies,
the dominant effect comes from the uncertainties in the strain amplitude calibration, as in~\cite{snsearch}. 
These vary from a few percent at lower frequencies to 10\% at higher frequencies in both L1 and H1.
For this paper, these uncertainties are conservatively set to 10\% for H1 and L1 at the times of the five CCSNe studied~\cite{c3,VIRGO}.
The error analysis leads to decreasing the detection efficiencies by 9.1\%.

For the waveforms coming from 2D simulations,
marginalizing over all unknown angles the waveform amplitude is reduced by a factor $\sqrt{5/18}$ that we apply to the efficiencies.
For optimally oriented CCSNe, the distance ranges for these models will be $\sqrt{18/5}$ times larger.

\begin{table*}[t] 
\caption{
  Waveforms from phenomenological and \textit{ad hoc} emission
  models described in the text.
  For each waveform, we give the
  emission type, journal reference, waveform identifier,
  angle-averaged root-sum-squared strain $h_\mathrm{rss}$, 
  the frequency $f_\mathrm{peak}$ at which the GW energy spectrum peaks, the
  emitted GW energy $E_\mathrm{GW}$, and available polarizations.
  See \cite{gossan:16,szczepanczykdcc:15} for details.
  As sine-Gaussian waveforms are \textit{ad hoc}, they can be 
  rescaled arbitrarily and do not have a defined 
  physical distance or $E_\mathrm{GW}$ value.
  } 
\begin{tabular}{clcrcc}
\hline
\hline
\multicolumn{1}{c}{Emission}
&\multicolumn{1}{c}{Waveform}
&\multicolumn{1}{c}{$\hrss$}
&\multicolumn{1}{c}{$f_\mathrm{peak}$}
&\multicolumn{1}{c}{$E_\mathrm{GW}$}
&\multicolumn{1}{c}{Polarizations}\\
Type 
&\multicolumn{1}{c}{Identifier}
&\multicolumn{1}{c}{$[10^{-22}\sqrt{\mathrm{s}}\,@ 10\,\mathrm{kpc}]$}
&\multicolumn{1}{c}{$[\mathrm{Hz}]$}
&\multicolumn{1}{c}{$[M_{\odot} c^{2}]$}\\
\hline
\hline
\multirow{6}{*}{\makecell{Long-lasting\\ Bar Mode~\cite{ott:10dcc}}}
&lb1-M0.2L60R10f400t100 &\phantom{0}1.480 &800 &\phantom{0}$2.984\times10^{-4}$ &$+,\times$\\
&lb2-M0.2L60R10f400t1000&\phantom{0}4.682 &800 &\phantom{0}$2.979\times10^{-3}$ &$+,\times$\\
&lb3-M0.2L60R10f800t100 &\phantom{0}5.920 &1600&\phantom{0}$1.902\times10^{-2}$ &$+,\times$\\
&lb4-M1.0L60R10f400t100 &\phantom{0}7.398 &800 &\phantom{0}$7.459\times10^{-3}$ &$+,\times$\\
&lb5-M1.0L60R10f400t1000&23.411 &800 &\phantom{0}$7.448\times10^{-2}$ &$+,\times$\\
&lb6-M1.0L60R10f800t25 &14.777 &1601 &\phantom{0}$1.184\times10^{-1}$ &$+,\times$\\
\hline
\multirow{4}{*}{\makecell{Torus Fragmentation\\ Instability~\cite{piro:07}}}
&piro1-M5.0$\eta$0.3 &\phantom{0}2.550 &2035 &\phantom{0}$6.773\times10^{-4}$ &$+,\times$\\
&piro2-M5.0$\eta$0.6 &\phantom{0}9.936 &1987 &\phantom{0}$1.027\times10^{-2}$ &$+,\times$\\
&piro3-M10.0$\eta$0.3&\phantom{0}7.208 &2033 &\phantom{0}$4.988\times10^{-3}$ &$+,\times$\\
&piro4-M10.0$\eta$0.6&28.084 &2041 &\phantom{0}$7.450\times10^{-2}$ &$+,\times$\\
\hline
\multirow{4}{*}{\makecell{sine-Gaussian~\cite{abadie:12s6burst}}}
&sg1-235HzQ8d9linear &--- &235 &--- &$+$\\
&sg2-1304HzQ8d9linear&--- &1304&--- &$+$\\
&sg3-235HzQ8d9elliptical &--- &235 &--- &$+,\times$\\
&sg4-1304HzQ8d9elliptical&--- &1304&--- &$+,\times$\\
\hline
\hline
\end{tabular}
\label{tab:models2}
\end{table*}

\subsubsection{Waveforms from multidimensional CCSN simulations}

The main mechanism behind a CCSN explosion is not yet fully understood and a complete review 
of the current state can be found in~\cite{janka:12a,1501.01688,kotake:07a,Foglizzo2015} and in references therein.
We divide the waveforms from multidimensional CCSN simulations
into two sets according to their explosion mechanisms.
In the first set, we consider a
\textit{neutrino-driven} explosion mechanism for non- or slowly-rotating progenitor stars.
We employ three waveform families: M\"uller~\cite{mueller:e12}, Ott~\cite{ott:13a} and Yakunin~\cite{yakunin:15}.
In this scenario, neutrino heating plays a crucial role in creating the explosion.
During the prompt convection, in the initial stages post bounce, 
GWs are emitted in the frequency range from 100-300\,Hz, while at later times, 
GWs up to around 2\,kHz can be expected~\cite{cerddurn:13,morozova:18}.
A typical duration for a GW transient is 0.5-1\,s~\cite{bethe:90,janka:12a,burrows:13a}.
The second set of waveforms, with simulations from the Scheidegger~\cite{scheidegger:10b} and Dimmelmeier~\cite{dimmelmeier:08} 
groups, utilizes the \textit{magnetohydrodynamically-driven (MHD-driven)} explosion mechanism 
for rapidly rotating progenitor stars.
The magnetic effects related to the rapid rotation may play a dominant role in creating the MHD-driven explosions.
Note, however, almost all (99\%)~\cite{Woosley2006, janka:12a} of explosions are believed to come from slowly rotating progenitor stars.

\textit{M\"uller et al}~\cite{mueller:e12} performed 3D 
simulations with a zero age main sequence (ZAMS) progenitor star of mass
$15\,M_\odot$ (L15-3 and W15-4), and a $20\,M_\odot$ (N20-2),
which we refer to as mul1, mul2 and mul3 respectively.
The simulations are three-dimensional and thus result in two polarizations.
The convective movement of infalling matter and its interaction with
the outer layers of the proto-neutron star result in GW emission 
in the frequency range 100-500\,Hz.

\textit{Ott et al}~\cite{ott:13a} produced a 3D simulation
with a $27\,M_\odot$ ZAMS progenitor star (ott1).
The explosion becomes aspherical due to strong convection while the SASI visibility is weak. This model is rotating and a strong 
burst of GWs appears at the beginning of the explosion.

\textit{Yakunin et al}~\cite{yakunin:15} delivers waveforms from 
four 2D simulations (providing only one polarization state)
corresponding to  $12\,M_\odot$, $15\,M_\odot$, $20\,M_\odot$, $25\,M_\odot$ 
ZAMS progenitor stars. We denote them as yak1, yak2, yak3 and yak4 respectively.
These waveforms capture several stages of the explosion. They show both low 
(SASI/convection) and high (g-mode) frequency components in their signals. 
Due to axisymmetry, the strain grows artificially over time, resulting in 
higher GW amplitudes than the 3D neutrino driven models.

\textit{Scheidegger et al}~\cite{scheidegger:10b} considers
effects on the GW signature due to the equation of state, 
the initial rotation rate, and the magnetic fields.
From an extensive set of waveforms, we extract three models, R1E1CA$_L$, 
R3E1AC$_L$, and R4E1FC$_L$, which we refer to as sch1, sch2, and sch3, respectively.  
All of these models are derived from the explosion of a $15\,M_\odot$ ZAMS progenitor star. 
The models are three dimensional and produce two GW polarization states. The degree of rotation 
varies between the models; model R1E1CA$_L$ has no rotation, which results in much lower GW 
energy in comparison to the rotating R3E1AC$_L$ and R4E1FC$_L$ models. 

\textit{Dimmelmeier et al}~\cite{dimmelmeier:08} performed 2D simulations (providing
linearly polarized waveforms) with a $15\,M_\odot$ ZAMS progenitor star.
The waveforms contain very strong 
GW emission at the initial core-collapse and bounce that lasts less than 20\,ms.
We employ three waveforms with various degrees of rotation from moderate to rapid (dim1-dim3).

\subsubsection{Extreme emission models}
\label{subsubsec:extreme}

Along with the more realistic simulated CCSN explosions, we also consider two extreme scenarios: 
the \textit{Long-Lasting Bar Mode}~\cite{ott:10dcc}, and the \textit{Torus Fragmentation Instability}~\cite{piro:07}.
The same models were used in~\cite{snsearch}, because even if 
they are unlikely to occur, they are not excluded~\cite{ETdesign:11}.

In the first scenario, a very rapidly rotating progenitor star induces a bar 
mode instability in the proto-neutron 
star~\cite{lai:95,brown:01,shibata:05,rmr:98,rotinst:05,ott:07prl,scheidegger:10b}.
This leads to large amplitude GWs that depend on the properties of the deformed proto-neutron star. 
We use the simple phenomenological bar model described in~\cite{ott:10dcc}.
In this model, we use the following parameterization:
the mass involved in the long-lasting bar mode of the proto-neutron star $M=\{0.2, 1.0\}\,M_\odot$, 
the radius $R=10$\,km and length $L=60$\,km of the bar, 
the spin frequency $f=\{400, 800\}$\,Hz along the direction perpendicular to the bar,
and the duration $t=\{25, 100, 1000\}$\,ms of the deformation.
We consider six waveforms, denoted as lb1-lb6, see Table~\ref{tab:models2} for more details.
Since the waveform amplitude is proportional to $M(L^2-3R^2)$ (see~\cite{ott:10dcc}),
any combination of $M$, $L$ and $R$ giving the same value for $M(L^2-3R^2)$ as 
the six waveforms, will produce waveforms identical to the lb1-lb6 waveforms. 
Therefore results for lb1-lb6 waveforms are a good representation of the broader sections of parameter space.

In the second scenario, Piro and Pfahl~\cite{piro:07} propose that, 
if a black hole and an accretion disk are formed during the collapse, 
the disk could fragment and large self-gravitating clumps of matter 
falling into the black hole would produce large amplitude GWs under 
the appropriate conditions. To model this signal we employ 
a simplified model~\cite{santamaria:11dcc} that depends on the mass of the 
central black hole $M_\mathrm{BH}=\{5,10\}\,M_\odot$ and the properties of the disk, 
namely the thickness of the torus  $\eta=\{0.3,0.6\}$ and the alpha 
viscosity parameter $\alpha=0.1$. The torus thickness is defined 
as $\eta=H/r$, where $H$ is the disk scale-height and $r$ the local radius.
For the disk model considered in~\cite{piro:07}, the mass of the fragmented clump 
is $M_f=0.53 \eta^3 M_\mathrm{BH}$. The GW amplitude is proportional to the reduced 
mass of the BH-clump system, $\mu=M_\mathrm{BH} M_f / (M_\mathrm{BH} + M_f)$, 
which for the parameter space considered here ($M_f \ll M_\mathrm{BH}$) is $\mu \approx M_f$.

\subsubsection{\textit{Ad-hoc} waveforms}

We employ \textit{ad hoc} waveforms to estimate the search sensitivity to short 
duration monochromatic signals that model GW emission in different frequency bands.
We use sine-Gaussian signals with a fixed central frequency $f_0=\{235,1304\}$\,Hz and duration 
$\tau=Q/(\sqrt{2}\pi f_0)$ where $Q=8.9$ is the quality factor. 
In our analysis, we use four \textit{ad hoc} waveforms denoted as sg1-sg4 that are linearly and elliptically 
polarized, see Table~\ref{tab:models2}.

\begin{figure}[hbt] 
\centering 
\includegraphics[width=1.0\columnwidth]{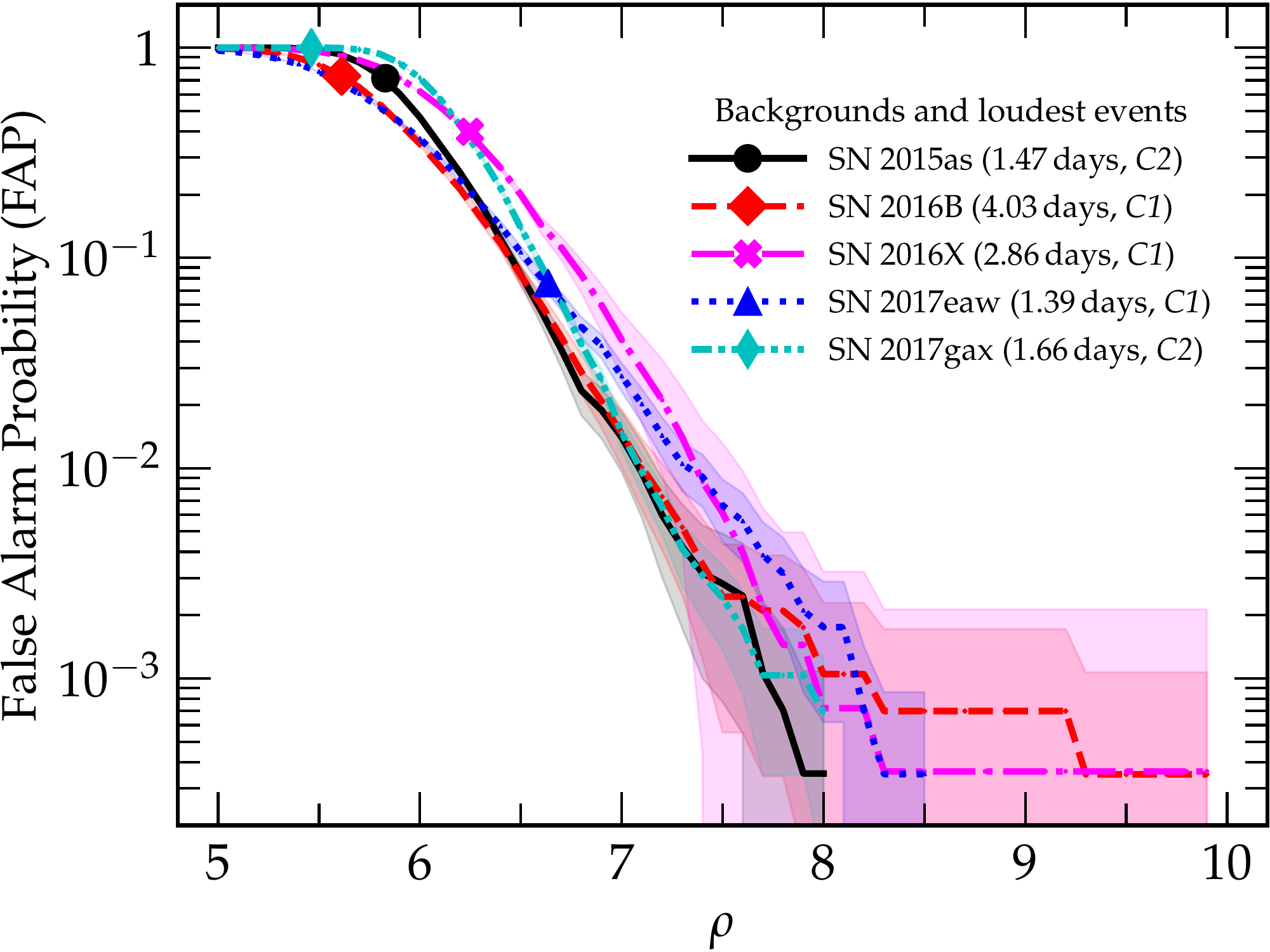} 
\caption{\label{fig:bkg}
    The False Alarm Probability (FAP) of background events
    together with the loudest events for each CCSN.
    The non-negligible values of the FAP indicate that all the loudest events appear compatible with the noise background.
    The shaded region is the $1\sigma$ error.
    The numbers in the parentheses are durations of the on-source 
    windows and the class where the loudest events belong.
  } 
\end{figure}

\section{Search Results}
\label{sec:results}

\begin{table}[htb]
\centering
\caption{List of the loudest events for each CCSN.
False alarm rate (FAR) and False Alarm Probability (FAP) for each of them
indicate that they are consistent with background noise.
}
\begin{tabular}{ll@{\hspace*{1em}}l@{\hspace*{1em}}l@{\hspace*{1em}}l}
	\hline
	\hline
	\multicolumn{1}{l}{Supernova} & \multicolumn{1}{c}{Class} & \multicolumn{1}{l}{\,$\rho$} & \multicolumn{1}{l}{{FAR [Hz]}} & \multicolumn{1}{c}{{FAP}} \\
	\hline
	\hline
	  SN 2015as  & \phantom{0}\phantom{0}\textit{C2} & 5.8 & 2.9\,$\times10^{-5}$ & 0.716 \\
	  SN 2016B   & \phantom{0}\phantom{0}\textit{C1} & 5.6 & 1.1\,$\times10^{-5}$ & 0.732 \\
	  SN 2016X   & \phantom{0}\phantom{0}\textit{C1} & 6.2 & 1.4\,$\times10^{-5}$ & 0.398 \\
	  SN 2017eaw & \phantom{0}\phantom{0}\textit{C1} & 6.6 & 1.3\,$\times10^{-6}$ & 0.076 \\
	  SN 2017gax & \phantom{0}\phantom{0}\textit{C2} & 5.5 & 9.7\,$\times10^{-5}$ & 1.000 \\
	\hline
	\hline
\end{tabular}
\label{tab:loudest}
\end{table}

Figure~\ref{fig:bkg} presents the background as a function of $\rho$ the coherent network signal-to-noise ratio
for all CCSNe. We plot the loudest events found in the OSWs
with further detailed information given in Table~\ref{tab:loudest}.
The non-negligible values of the false alarm probabilities indicate that all the results appear compatible with the noise background.
For each CCSN source, we estimate the search efficiency using the waveforms described in Sec.~\ref{sec:sim}
considering events with a rank value smaller than the loudest event FAR.

\subsection{Detection efficiency vs.~distance}
\label{sec:eff}

\begin{figure*}[t] 
  \begin{minipage}[c][][t]{0.495\textwidth}
    \vspace*{\fill}
    \flushleft
    \includegraphics[width=0.96\linewidth]{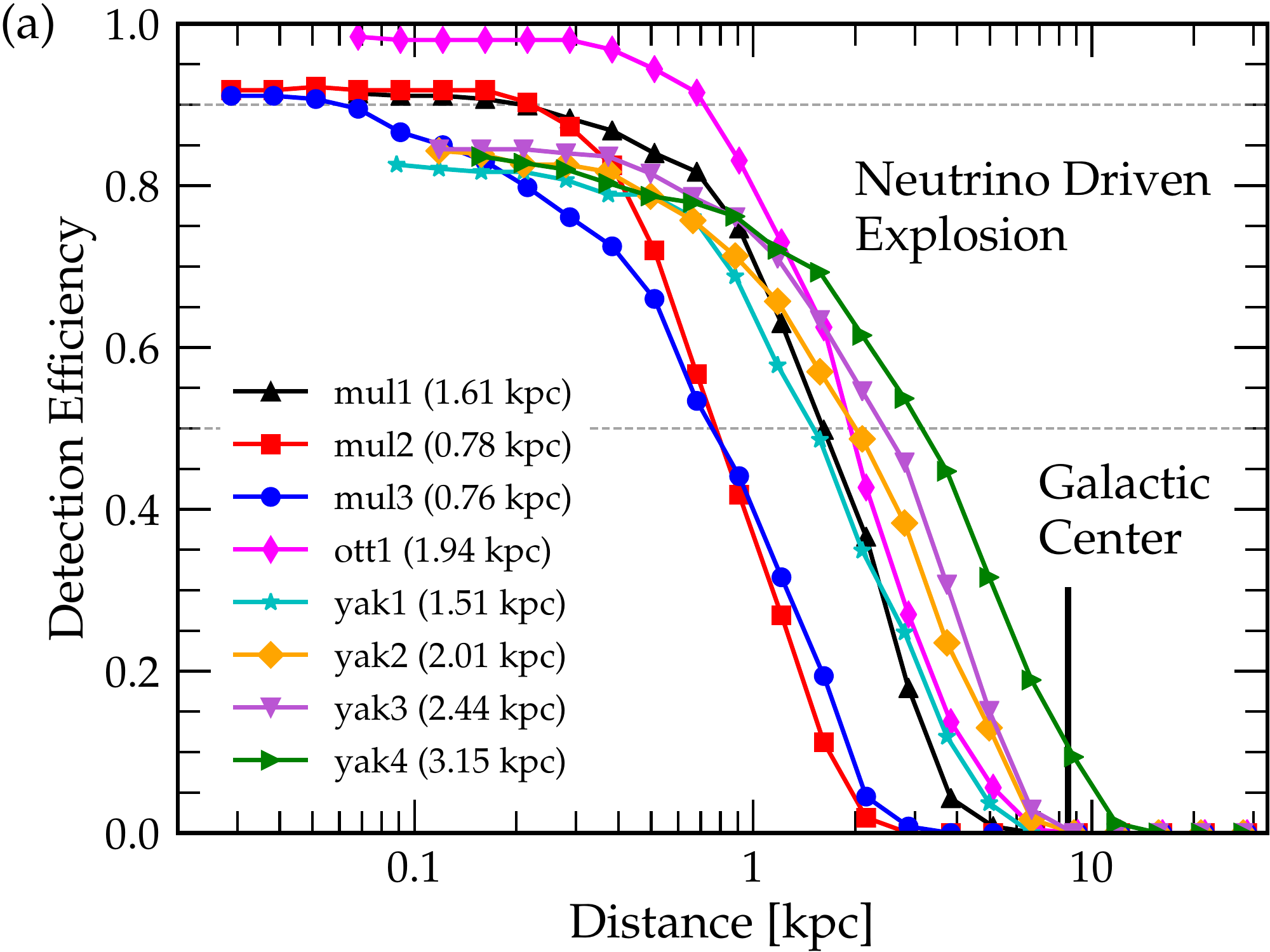}
    \includegraphics[width=0.96\linewidth]{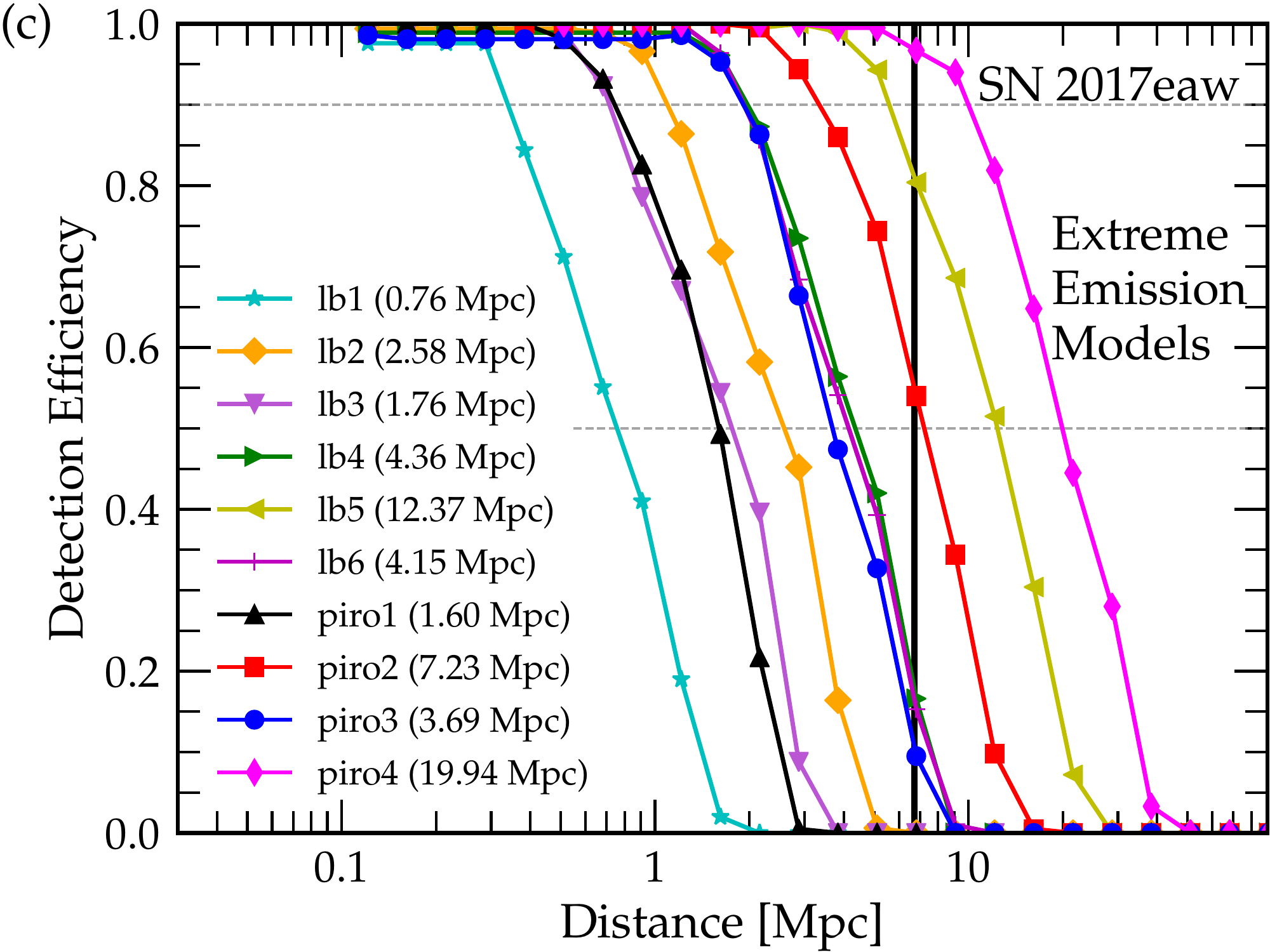}
  \end{minipage}%
  \begin{minipage}[c][][t]{0.495\textwidth}
    \vspace*{\fill}
    \flushright
    \includegraphics[width=0.96\linewidth]{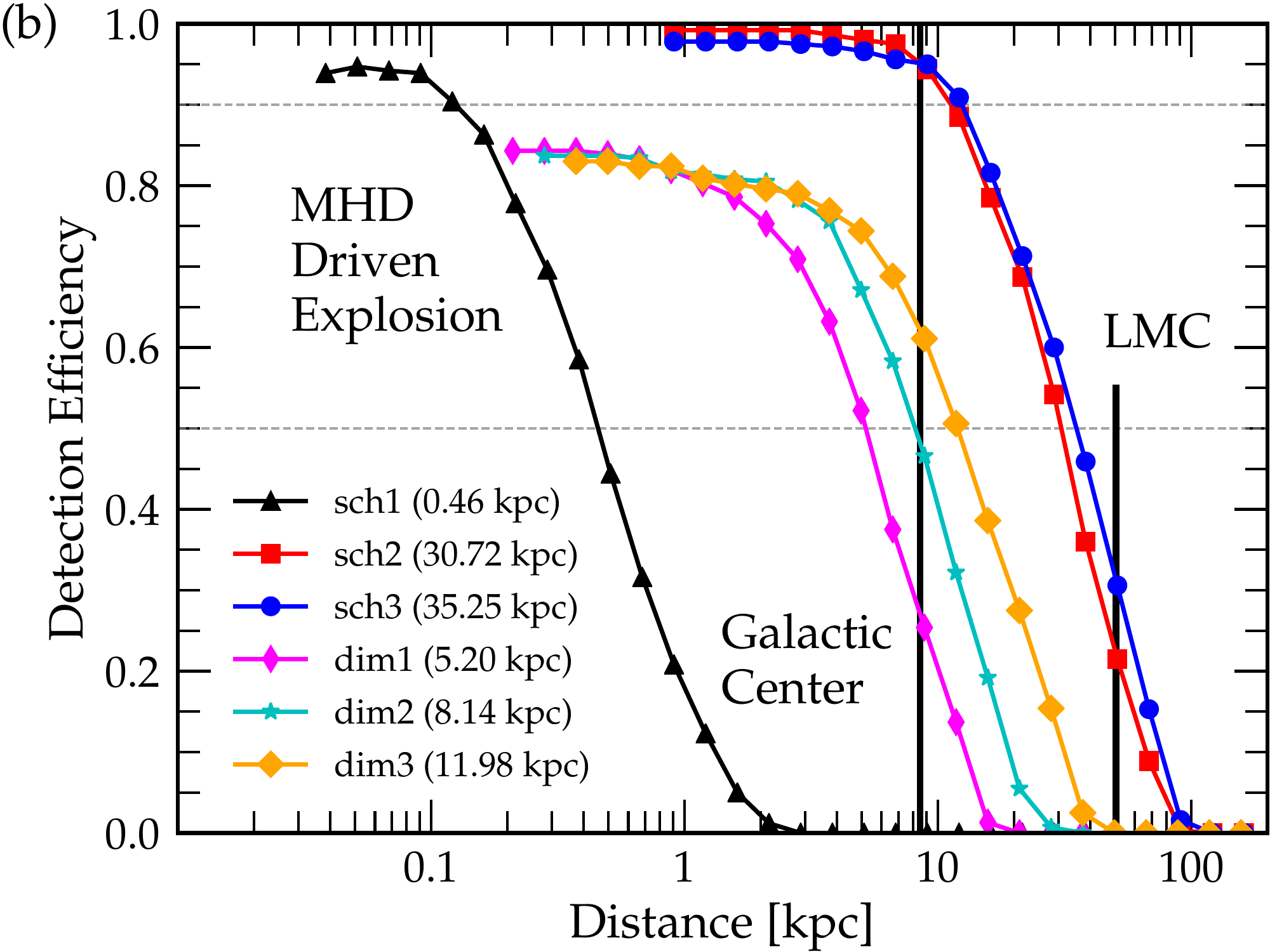}
    \includegraphics[width=0.96\linewidth]{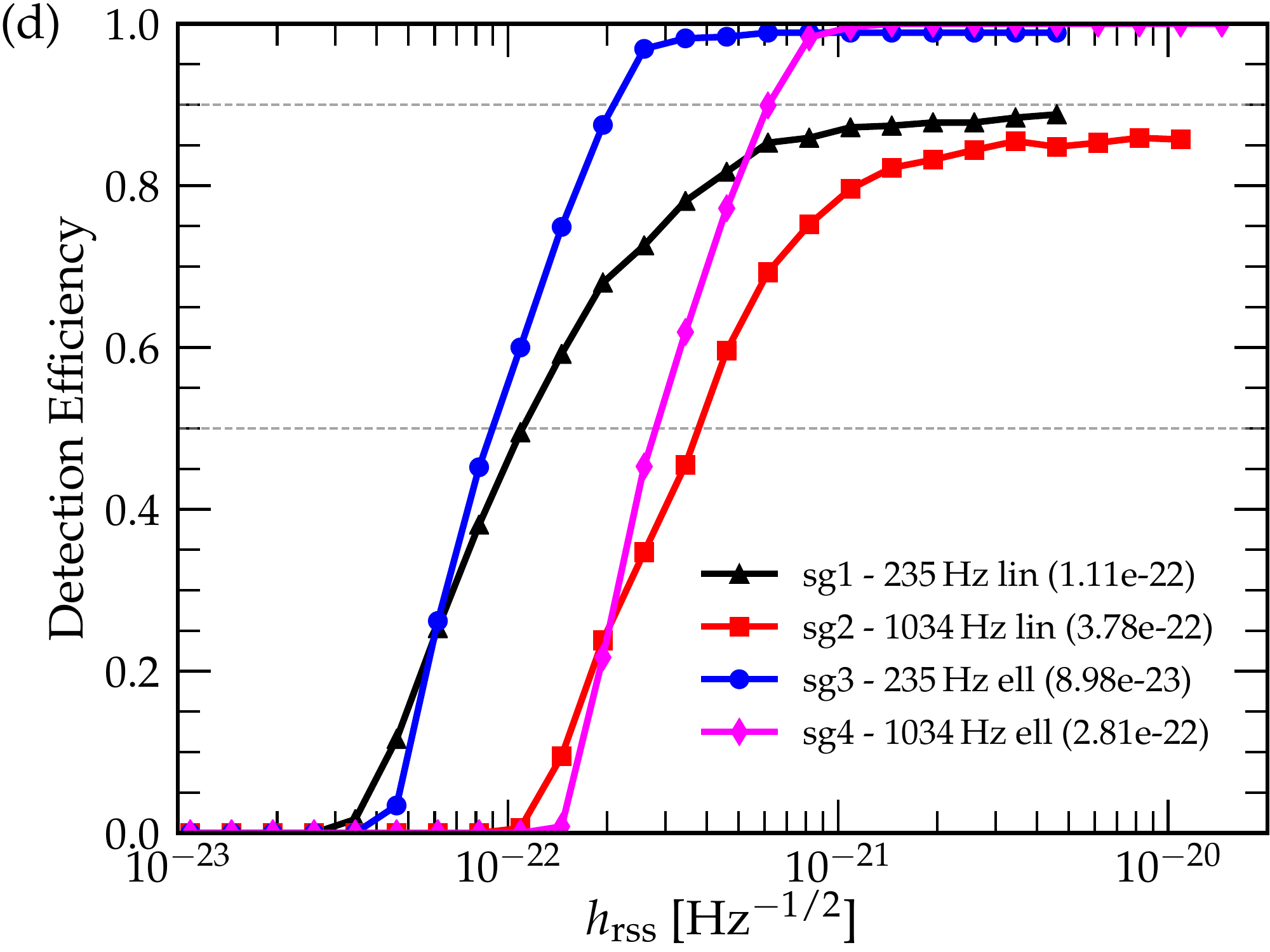}
  \end{minipage}
  \caption{\label{fig:sn2016B}
     Panels (a), (b), and (c) show the efficiency
     as a function of distance using 3 families of waveforms for a source located at the position and time of SN~2017eaw.
     Panel (d) provides the detection efficiency for \textit{ad hoc} sine-Gaussian waveforms as a function of $\hrss$, which we use to constrain GW energy and discuss in Sec.~\ref{sec:egw}.
     The numbers in parentheses for the models plotted in (a), (b), and (c)
     are the distances at which the detection efficiency equals 50\%.
     For (d), the numbers in parentheses are the $\hrss$ values resulting in 50\% detection efficiencies.
     The detection reach for neutrino-driven explosions is limited to a few kpc while
     for magnetorotationally-driven (MHD-driven) explosions it covers the Milky Way and 
     the detection efficiency at the distance of the Large Magellanic Cloud (LMC), that hosted SN~1987A, is non-zero.
     Further discussion can be found in Sec.~\ref{sec:eff}.
     The distance reaches for extreme emission models in (c) exceed the distance
     of SN~2017eaw. Given that there was no GW detection, we are able to exclude some
     of the parameter spaces for these models, which we discuss in Sec.~\ref{sec:mep}.
     The dashed lines show $50\%$ and $90\%$ detection efficiencies.
     }
  \label{fig:eff}
\end{figure*}

\begin{table*}[hbt]
\caption{Distance reaches (in kpc), defined as the distance with a 50\% detection efficiency,
for neutrino-driven explosions from M\"uller (mul1-mul3), Ott (ott1) and Yakunin (yak1-yak4), as well as
MHD-driven explosions from Scheidegger (sch1-sch3) and Dimmelmeier (dim1-dim3).
\label{tab:dist_nu}}
\begin{center}
\begin{tabular}{p{2cm}|C{0.8cm}C{0.8cm}C{0.8cm}C{0.8cm}C{0.8cm}C{0.8cm}C{0.8cm}C{0.8cm}|C{0.8cm}C{0.8cm}C{0.8cm}C{0.8cm}C{0.8cm}C{0.8cm}}
\hline
\hline 
Supernova & mul1 & mul2 & mul3 & ott1 & yak1 & yak2 & yak3 & yak4 & sch1 & sch2 & sch3 & dim1 & dim2 & dim3 \\
\hline
\hline
SN 2015as & 1.53 & 0.84 & 0.97 & 2.20 & 1.73 & 1.89 & 2.05 & 2.47 & 0.49 & 34.76 & 41.32 & 5.57 & 7.34 & 14.06 \\
SN 2016B & 1.32 & 0.70 & 0.58 & 1.74 & 0.65 & 1.07 & 1.32 & 2.12 & 0.40 & 26.66 & 34.88 & 3.82 & 5.24 & 12.09 \\
SN 2016X & 1.26 & 0.57 & 0.66 & 1.72 & 0.50 & 0.84 & 0.81 & 1.62 & 0.37 & 26.02 & 32.13 & 2.52 & 3.80 & 10.22 \\
SN 2017eaw & 1.61 & 0.78 & 0.76 & 1.94 & 1.51 & 2.01 & 2.44 & 3.15 & 0.46 & 30.72 & 35.25 & 5.20 & 8.14 & 11.98 \\
SN 2017gax & 2.40 & 1.15 & 1.24 & 3.09 & 2.30 & 2.75 & 2.95 & 5.08 & 0.65 & 42.29 & 53.55 & 8.04 & 10.19 & 22.35 \\
\hline
\hline
\end{tabular} 
\end{center}
\end{table*}

We provide detection efficiencies for waveforms listed in 
Table~\ref{tab:models1} and Table~\ref{tab:models2}.
Figure~\ref{fig:eff} presents the detection efficiencies
for SN~2017eaw, the closest CCSN in this search.
For reference, the plots show the distances to the Galactic center (8.5\,kpc),
the Large Magellanic Cloud (49.6\,kpc~\cite{Pietrzyski2019}) that hosted SN~1987A,
and the distance to the host galaxy of SN~2017eaw, NGC~6946.
For each model, we determine the distance corresponding to a 50\% detection efficiency.
Distance reaches for each CCSN for neutrino-driven explosions and MHD-driven explosions are summarized in Table~\ref{tab:dist_nu}
and extreme emission models in Table~\ref{tab:dist_extreme}.
For each model the distances are consistent across CCSNe
and these distances are around 3-5 times further than in~\cite{snsearch}.
The largest distances are obtained for SN~2017gax. This can be explained by the fact that the 
loudest event for this CCSN has the lowest $\rho$ value and the network sensitivity 
at the time of the CCSN was better over the duration of the OSW.

\begin{table*}
\caption{Distance reaches (in Mpc), defined as the distance with a 50\% detection efficiency,
for extreme emission models Torus Fragmentation Instability (piro1-piro4) and
Long-­lasting Bar Mode (lb1-lb6).
\label{tab:dist_extreme}}
\begin{center}
\begin{tabular}{p{2cm}|C{0.8cm}C{0.8cm}C{0.8cm}C{0.8cm}C{0.8cm}C{0.8cm}C{0.8cm}C{0.8cm}C{0.8cm}C{0.8cm}}
\hline
\hline 
Supernova & piro1 & piro2 & piro3 & piro4 & lb1 & lb2 & lb3 & lb4 & lb5 & lb6 \\
\hline
\hline
SN 2015as & 1.33 & 7.13 & 3.83 & 19.70 & 0.93 & 2.93 & 1.92 & 4.59 & 15.24 & 4.86 \\
SN 2016B & 1.31 & 7.04 & 3.47 & 17.94 & 0.80 & 2.64 & 1.80 & 4.24 & 13.69 & 4.50 \\
SN 2016X & 1.32 & 6.86 & 3.20 & 19.55 & 0.73 & 2.36 & 1.46 & 3.83 & 12.16 & 3.73 \\
SN 2017eaw & 1.60 & 7.23 & 3.69 & 19.94 & 0.76 & 2.58 & 1.76 & 4.36 & 12.37 & 4.15 \\
SN 2017gax & 1.81 & 10.04 & 5.22 & 27.79 & 1.23 & 3.55 & 2.63 & 6.16 & 19.03 & 6.40 \\
\hline
\hline
\end{tabular} 
\end{center}
\end{table*}

For the neutrino-driven explosions summarized in Table~\ref{tab:dist_nu} and the upper left panel of
Figure~\ref{fig:eff}, the detection distance reached less than 5~kpc.
None of these models reach the Galactic center,
however a few of the waveforms have non-zero detection probabilities at that distance.
The least detectable models are the M\"uller waveforms because they are the least energetic.
The most detectable models are the Yakunin waveforms and the reach increases
with progenitor mass.
The Ott model has a smaller detection reach compared to the Yakunin waveforms,
but also has higher detection efficiency at small distances.

\begin{table*}
\caption{Gravitational-wave energy emission constraints at
  50\% detection efficiency.
  We assumed isotropic GW emission for the four \textit{ad hoc} waveforms.
  The most stringent constraints (in bold) are achieved for the closest event, SN~2017eaw (6.72\,Mpc).
  They are two orders of magnitude smaller than the
  results obtained in Ref.~\cite{snsearch}.
  The energies obtained for 235\,Hz emission are comparable to the
  typical explosion energy ($\sim10^{51}$\,erg) and 
  a typical kinetic energy of CCSN ejecta ($\sim10^{51}$\,erg).
\label{tab:energy}}
\begin{center}
\begin{tabular}{c|c|c|c|c|c}
\hline
\hline 
Supernova & Quantity & sg1 - 235\,Hz lin & sg2 - 1304\,Hz lin & sg3 - 235\,Hz ell & sg4 - 1304\,Hz ell \\
\hline
\hline
\multirow{3}{*}{SN 2015as}
& $\hrss$\,[\hrssu] & 8.96e-23 & 2.95e-22 & 7.72e-23 & 2.58e-22 \\ \cline{2-6}
& $E_\mathrm{GW}\,[\mathrm{erg}]$ & 6.20e+51 & 2.07e+54 & 4.60e+51 & 1.58e+54 \\ \cline{2-6}
& $E_\mathrm{GW}\,[M_{\odot}c^2]$ & 3.47e-03 & 1.16e+00 & 2.57e-03 & 8.82e-01 \\
\hline
\hline
\multirow{3}{*}{SN 2016B}
& $\hrss$\,[\hrssu] & 1.15e-22 & 4.07e-22 & 8.72e-23 & 2.71e-22 \\ \cline{2-6}
& $E_\mathrm{GW}\,[\mathrm{erg}]$ & 9.59e+51 & 3.70e+54 & 5.51e+51 & 1.64e+54 \\ \cline{2-6}
& $E_\mathrm{GW}\,[M_{\odot}c^2]$ & 5.37e-03 & 2.07e+00 & 3.08e-03 & 9.15e-01 \\
\hline
\hline
\multirow{3}{*}{SN 2016X}
& $\hrss$\,[\hrssu] & 1.33e-22 & 4.52e-22 & 9.86e-23 & 3.11e-22 \\ \cline{2-6}
& $E_\mathrm{GW}\,[\mathrm{erg}]$ & 8.60e+51 & 3.05e+54 & 4.70e+51 & 1.44e+54 \\ \cline{2-6}
& $E_\mathrm{GW}\,[M_{\odot}c^2]$ & 4.81e-03 & 1.71e+00 & 2.63e-03 & 8.08e-01 \\
\hline
\hline
\multirow{3}{*}{SN 2017eaw}
& $\hrss$\,[\hrssu] & 1.11e-22 & 3.78e-22 & 8.98e-23 & 2.81e-22 \\ \cline{2-6}
& $E_\mathrm{GW}\,[\mathrm{erg}]$ & \textbf{1.16e+51} & \textbf{4.17e+53} & \textbf{7.63e+50} & \textbf{2.30e+53} \\ \cline{2-6}
& $E_\mathrm{GW}\,[M_{\odot}c^2]$ & \textbf{6.49e-04} & \textbf{2.33e-01} & \textbf{4.27e-04} & \textbf{1.28e-01} \\
\hline
\hline
\multirow{3}{*}{SN 2017gax}
& $\hrss$\,[\hrssu] & 6.80e-23 & 2.35e-22 & 5.72e-23 & 1.98e-22 \\ \cline{2-6}
& $E_\mathrm{GW}\,[\mathrm{erg}]$ & 3.76e+51 & 1.38e+54 & 2.66e+51 & 9.79e+53 \\ \cline{2-6}
& $E_\mathrm{GW}\,[M_{\odot}c^2]$ & 2.10e-03 & 7.71e-01 & 1.49e-03 & 5.47e-01 \\
\hline
\hline
\end{tabular} 
\label{tab:egw}
\end{center}
\end{table*}

Table~\ref{tab:dist_nu} and the upper right panel of Figure~\ref{fig:eff}
also present a summary of distance reaches for MHD-driven explosions. 
The distance reaches for most of these models are an order of magnitude larger 
than for the neutrino-driven explosions.
Some MHD-driven explosion models reach to the distance of Large Magellanic Cloud.
If a MHD-driven supernova were to explode at the distance of SN~1987A, around 50\,kpc away,
we have a non-zero chance of detecting it.
The detectable range for sch1 is two orders of magnitude shorter compared to those of sch2 and sch3.
This Scheidegger model has a lower amplitude due to its slower rotation.

Distance reaches for the extreme emission models are given in Table~\ref{tab:dist_extreme}
and depicted in the bottom left panel of Figure~\ref{fig:eff}.
The ranges are on the order of several Mpc up to nearly 28\,Mpc for the most extreme model.
The reaches of a few waveforms exceed the distance of SN~2017eaw.
Given the null detections, this means we can begin to exclude these models as discussed in Sec.~\ref{sec:mep}.

The detection efficiencies for the linearly polarized waveforms (Dimmelmeier and Yakunin) do not reach 
unity even at small distances because the network of detectors is not sensitive to both polarizations
for any sky position at a given time. There are sky positions where the detectors are insensitive 
to one of the polarizations and even a large amplitude signal with only one polarization will not be detectable. 
Waveforms with two polarizations are more efficiently detected than linearly polarized signals.

\subsection{Constraints on GW energy emission}
\label{sec:egw}

Similarly to~\cite{snsearch} we provide constraints on the GW energy emission from CCSNe.
This is the minimum energy emitted in GWs needed to be detectable with 50\% probability.
We calculate these constraints individually for each CCSN.
We probe low- and high-frequency GW emission using sine-Gaussian \textit{ad hoc} waveforms with
central frequencies of 235\,Hz and 1304\,Hz (see Table~\ref{tab:models2}).
These waveforms do not have physical meaning,
so we plot detection efficiency as a function of $\hrss$
(eqn.~\eqref{eq:hrss}) instead of distance.
We assume isotropic emission with a total energy of
\begin{equation}
\label{eqn:energy}
E_\mathrm{GW} = \frac{\pi^2 c^3}{G}D^2 f^2_0\hrss^2 \, ,
\end{equation}
where $f_0$ is the peak GW frequency of the sine-Gaussian, and $D$ is the distance to the source~\cite{sutton:13}.

The bottom right panel of Figure~\ref{fig:eff} shows the detection efficiency
versus $\hrss$ for the four \textit{ad hoc} waveforms. 
The quoted $\hrss$ for each sine-Gaussian corresponds to the strain with a 
50\% detection efficiency. The best sensitivity is
achieved for the sine-Gaussians around 235\,Hz (sg1 and sg3), which is a result of the
lower noise level of the detectors at this frequency. 
The efficiency curves of the elliptically polarized 
waveforms (sg3 and sg4) flatten at higher detection efficiencies
compared to the efficiencies for linearly polarized waveforms (sg1 and sg2) for the reason discussed in Sec.~\ref{sec:eff}.

In Table~\ref{tab:egw}, we report GW energy constraints for each CCSN.
For the \textit{ad hoc} waveforms with peak frequency at 235\,Hz (sg1 and sg3),
the GW energy constraints are consistently on the order of $10^{-3}\,M_\odot c^2$ or less.
The lowest achieved energy constraints are obtained
for SN~2017eaw at low frequency, 
$4.27\times 10^{-4}\,M_\odot c^2$ ($7.63\times 10^{50}$\,erg), 
and high frequency, $1.28\times 10^{-1}\,M_\odot c^2$ ($2.30\times 10^{53}$\,erg).
For both low and high frequency emission, the energy constraints
are two orders of magnitude stronger than in the search with the initial interferometer data~\cite{snsearch}.
This improvement is due to the improved sensitivity of the detectors and the closer distance 
of SN~2017eaw (6.72\,Mpc) in comparison to SN~2007gr (10.55\,Mpc).
However, these energy constraints are still a few orders of magnitude larger than 
the energies predicted from multidimensional simulations (Table~\ref{tab:models1}), that
lie between around $10^{-11}\,M_\odot c^2$ and $10^{-7}\,M_\odot c^2$.

The GW energy constraints obtained in this search
can be compared to the energy budget of a CCSN.
The energy available during collapse is approximately the gravitational binding energy 
of the final neutron star remnant, which is typically 
$1.5\times 10^{-1}\,M_\odot c^2$ ($3\times 10^{53}$\,erg,~\cite{lattimer:01,Kotake2006,Lamers2017book}).
Around 99\% of that energy is radiated via neutrinos
during the cooling of the proto-neutron star~\cite{vissani:15} and
the remaining $\sim1\%$ is mainly transferred into kinetic energy. 
In a realistic scenario, only a small fraction of the explosion's energy is radiated
in the GW spectrum (Sec.~\ref{sec:sim}).

In Sec.~\ref{sec:sim} we describe several processes emitting GWs.
Some of them (e.g., SASI and convection) are related to the movement of
matter that is ejected during an explosion.
Again, according to the multidimensional simulations, only a small portion of this energy
is converted into GWs.
The kinetic energy of CCSN ejecta is typically on the order of $5.5\times 10^{-4}\,M_\odot c^2$
($10^{51}$\,erg)~\cite{Smartt2009,Tanaka2009,Utrobin2011}.
Specifically, estimates of the kinetic energy in the ejecta of SN~2015as and SN~2017eaw are 
$2.5\times 10^{-3}\,M_\odot c^2$ ($4.4\times 10^{51}$\,erg)~\cite{1802.06731} 
and $1.1\times 10^{-3}\,M_\odot c^2$ ($2.0\times 10^{51}$\,erg)~\cite{Tsvetkov2018}, respectively.
The current GW constraints at low frequencies are comparable with these values.
Specifically, the low frequency (235\,Hz) constraints for SN~2017eaw are roughly an 
order of magnitude below the kinetic energy of CCSN ejecta.

For extreme emission models, the GW energies are orders of magnitude
larger than those predicted for multidimensional simulations, as seen in Table~\ref{tab:models2}.
The energies of these extreme emission processes range from $2.98\times 10^{-4}\,M_\odot c^2$
up to even $1.18\times 10^{-1}\,M_\odot c^2$.
Our energy constraints are comparable with these values, 
but the comparison would not be correct as the \textit{ad hoc} signals 
and the extreme emission model waveforms frequency content are different.

\subsection{Model exclusion statements for extreme emission models}
\label{sec:mep}

Along with constraining the GW energy emitted by CCSNe,
we also constrain two models of extreme GW emission.
As described in Sec.~\ref{sec:eff}, 
for a few waveforms of the extreme emission models, 
the distance reaches
exceed the distances of the CCSNe analyzed in this search.
Given no GW detection, these models most likely do not describe 
correctly the CCSN explosion phenomena.
Similarly to Ref.~\cite{snsearch}, we consider a \textit{standard candle} approach,
that is we assume that each CCSN emits an identical GW signal.
To characterize the models, we use waveforms
that probe sample regions of the parameter spaces of these models
(see Table~\ref{tab:models2} and Sec.~\ref{subsubsec:extreme}).

The method for excluding models from multiple astrophysical sources
is described in detail in~\cite{kalmus:13}.
In this method we use the detection efficiency, $\mathcal{E}(d)$, which  
is a function of the distance, $d$. If a GW transient is strong and detectable but
arrives at the detectors when coincident data is not available,
then the model that predicts such a transient cannot be excluded.
Therefore, we need to take into account the 
coincident duty factor, $a\in[0,1]$.
We define the \textit{reduced detection efficiency} as 
\begin{equation}
 \epsilon(d) = a \times \mathcal{E}(d).
\end{equation}
Given no GW detection, the reduced detection efficiency can also be understood as a model exclusion probability.
For example, the detection efficiency for the piro4 waveform reaches 96.7\%
(see Figure~\ref{fig:eff}) at the distance of SN~2017eaw (6.72\,Mpc).
When we take into account the effect of the 48.8\% coincident duty factor 
for this CCSN 
(see Table~\ref{tab:sne}) the reduced detection efficiency is 47.2\%.
Hence we are confident with 47\% probability that the piro4 model does not
correctly describe the nature of a CCSN engine.

We then combine model exclusion probabilities obtained for each 
CCSN by multiplying the probabilities of not detecting a signal. 
The overall model exclusion probability
\be
P_\mathrm{excl} = 1 - \prod_{i =1}^N (1-\epsilon_i(d_i))\,, \label{eq:reach}
\ee  
where $N$ is the number of CCSNe.

\begin{table*}
\caption{Model Exclusion Probabilities ($P_\mathrm{excl}$) for extreme emission models
with a standard candle approach (see Sec.~\ref{sec:mep} for details of the method).
We infer that if bars are created generically in type-II and type-Ib/c supernovae, 
then the deformations are preferably small.
If central black holes are created in CCSN, 
then the accretion tori around them are either non-fragmented or rather thin.
\label{tab:mep}}
\begin{center}
\begin{tabular}{p{16mm}|p{19mm}|C{0.8cm}C{0.8cm}C{0.8cm}C{0.8cm}|C{0.8cm}C{0.8cm}C{0.8cm}C{0.8cm}C{0.8cm}C{0.8cm}}
\hline
\hline

\multicolumn{2}{c|}{Supernova} & piro1 & piro2 & piro3 & piro4 & lb1 & lb2 & lb3 & lb4 & lb5 & lb6 \\
\hline
\hline
\multirow{5}{*}{\makecell{Reduced\\ Detection\\ Efficiency\\ $\epsilon$ [\%]}}
& SN 2015as & 0.0 & 0.2 & 0.0 & 18.0 & 0.0 & 0.0 & 0.0 & 0.0 & 8.4 & 0.0 \\
& SN 2016B & 0.0 & 0.1 & 0.0 & 16.0 & 0.0 & 0.0 & 0.0 & 0.0 & 5.5 & 0.0 \\
& SN 2016X & 0.0 & 0.0 & 0.0 & 9.8 & 0.0 & 0.0 & 0.0 & 0.0 & 3.1 & 0.0 \\
& SN 2017eaw & 0.0 & 26.8 & 5.2 & 47.2 & 0.0 & 0.0 & 0.0 & 8.7 & 39.5 & 8.0 \\
& SN 2017gax & 0.0 & 0.2 & 0.0 & 48.7 & 0.0 & 0.0 & 0.0 & 0.0 & 28.6 & 0.0 \\
\hline
\hline
\multicolumn{2}{c|}{$P_\mathrm{excl}$ [\%]} & 0.0 & \textbf{27.2} & \textbf{5.2} & \textbf{83.2} & 0.0 & 0.0 & 0.0 & \textbf{8.7} & \textbf{63.8} & \textbf{8.0} \\
\hline
\hline
\end{tabular} 
\label{tab:mep}
\end{center}
\end{table*}

The results for the ten waveforms described in Sec.~\ref{subsubsec:extreme} are shown in Table~\ref{tab:mep}.
The greatest $P_\mathrm{excl}=83.2\%$ is obtained for the piro4 waveform
and the largest contributions come from SN~2017eaw and SN~2017gax because the detection ranges for these CCSNe are larger than their distances.
Although SN~2017eaw makes the most important contribution to the model exclusion statements,
the most energetic models, piro4 and lb5, are constrained by all CCSNe.

For the Torus Fragmentation Instability model, the waveforms are characterized
by the mass of a central black hole and the thickness of a torus around it.
The clump masses $M_f$ for the piro1-piro4 waveforms are
0.072$\,M_\odot$, 0.576$\,M_\odot$, 0.144$\,M_\odot$, 1.152$\,M_\odot$ respectively.
There is a correlation between the mass of the fragment and $P_\mathrm{excl}$ because the amplitude of the waveform scales approximately with $M_f$.
We conclude that
if central black holes are created in type-II and type-Ib/c supernovae after core-collapse,
then any clumps formed by fragmentation are preferably small ($M_f \lesssim 1 M_\odot$).
Moreover, if the torii are created around black holes, they are either non-fragmented or rather thin,
for the disk model considered in Ref.~\cite{piro:07}.

For the Long-Lasting Bar Mode model, the parameter space is larger
than for the Torus Fragmentation Instability model. Three models have non-zero $P_\mathrm{excl}$ values,
lb4, lb5, and lb6.
All three models have large values of $M(L^2-3R^2)=3300$\,$M_\odot$km$^2$, 
which corresponds either to $R\le10$\,km proto-neutron stars with large asymmetries 
($L/2R>2.5$) or large proto-neutron stars ($R\approx20$\,km) with moderate asymmetries ($L/2R\approx1.5-3.5$). 
If bars are created generically in type-II and type-Ib/c supernovae, the deformations are preferably small.
The largest $P_\mathrm{excl}$ among these models is obtained for lb5. This waveform lasts 1\,s,
while lb4 and lb6 are 100\,ms and 25\,ms respectively.
It seems that, for a proto-neutron star with $R\le10$\,km, if bars with strong 
deviations from axisymmetry are created ($L/2R>3$) in CCSNe, then they are rather 
short lived. Larger proto-neutron stars ($R\approx20$\,km) could still have
large deformations and be unobservable.

The constraint for lb5 limits the possible maximum deformations in type-II 
and type-Ib/c supernovae. If bars are created they are probably small. 
In case deformations are large the proto-neutron star is either very compact 
($R\sim5$\,km) or the bar is short lived ($<1000$\,ms). 
These results are consistent with the current theoretical understanding 
of bar-mode instabilities, which are expected to appear early after bounce when the 
proto-neutron star mass is relatively low ($M \le 1$\,$M_\odot$) and  its radius 
large ($R\ge 20$\,km). The amplitude and duration are also expected to be 
severely limited by the presence of strong magnetic fields, magneto-rotational 
turbulence, and shear instabilities.

\section{Summary and Discussion}
\label{sec:summaries}

We present the results of a search for GWs from CCSNe
with the Advanced LIGO and Advanced Virgo detectors
during the first and second observing runs (2015-2017).
Five CCSNe within 20\,Mpc are used
for the astrophysical statements:
SN~2015as, SN~2016B, SN~2016X, SN~2017eaw, and SN~2017gax.
We have not found any significant GW candidate.
All the loudest events are consistent with background events.

We provide the distance reaches at $50\%$ detection efficiencies for both realistic and extreme GW emission models.
For the neutrino driven explosions the distances do not exceed 5\,kpc, while the distance 
ranges for the magnetorotationally driven explosions reach 54\,kpc.
The distance reaches for extreme emission models
can be as large as 28\,Mpc, which exceed the distances of CCSNe analyzed in this search.
Given no GW detection, this gives us an opportunity to estimate the exclusion probabilities for the most extreme models.

We derive GW energy constraints for generic low and high frequency 
GW emissions at 235\,Hz and 1304\,Hz respectively
using linearly and elliptically polarized \textit{ad-hoc} sine-Gaussian waveforms.
The constraints are around $10^{-3}\,M_\odot c^2$ and
$10^{-1}\,M_\odot c^2$ for low and high frequency GW emission, respectively.
The best GW emission constraints we obtained are for SN~2017eaw
of $4.27\times 10^{-4}\,M_\odot c^2$ ($7.63\times 10^{50}$\,erg) for 
low frequency emission and $1.28\times 10^{-1}\,M_\odot c^2$ ($2.30\times 10^{53}$\,erg)
for high frequency emission. These are two orders of magnitude more stringent than in~\cite{snsearch},
but still a few orders of magnitude larger than predicted from multidimensional simulations.
The low frequency emission constraints are
comparable to the typical kinetic energy of CCSN ejecta.

We provide the first supernova model constraints
based on O1 and O2 data with a standard candle approach. The most extreme emission
models, piro4 and lb5, are constrained at the level of 83.2\% and 63.8\% respectively.
Out of ten waveforms, we place limits on six of them with 5\% to 83\% 
exclusion probabilities.
Based on our results, we conclude that if central black holes are created
in type-II and type-Ib/c supernovae, the sizes of the fragments 
are preferably small. 
Moreover, if disks around central black holes are created, then 
they are either non-fragmented or rather thin.
If bars are created, they are probably small.
In cases where deformations of the proto-neutron star are large, they are either very compact 
($R\sim5$\,km) or they shortly lived ($<1000$\,ms).

These  model exclusion statements are the first constraints
on CCSN engines based on GW data. In the future, 
with targeted searches and upgraded detectors (third observing run and beyond) 
it will be possible to further exclude the extreme emission models
and better constrain the GW energy emitted by CCSN engines, making both more
astrophysically meaningful.

\begin{acknowledgments}\label{sec:acknowledgments}

The authors gratefully acknowledge the support of the United States
National Science Foundation (NSF) for the construction and operation of the
LIGO Laboratory and Advanced LIGO as well as the Science and Technology Facilities Council (STFC) of the
United Kingdom, the Max-Planck-Society (MPS), and the State of
Niedersachsen/Germany for support of the construction of Advanced LIGO 
and construction and operation of the GEO600 detector. 
Additional support for Advanced LIGO was provided by the Australian Research Council.
The authors gratefully acknowledge the Italian Istituto Nazionale di Fisica Nucleare (INFN),  
the French Centre National de la Recherche Scientifique (CNRS) and
the Foundation for Fundamental Research on Matter supported by the Netherlands Organisation for Scientific Research, 
for the construction and operation of the Virgo detector
and the creation and support  of the EGO consortium. 
The authors also gratefully acknowledge research support from these agencies as well as by 
the Council of Scientific and Industrial Research of India, 
the Department of Science and Technology, India,
the Science \& Engineering Research Board (SERB), India,
the Ministry of Human Resource Development, India,
the Spanish  Agencia Estatal de Investigaci\'on,
the Vicepresid\`encia i Conselleria d'Innovaci\'o, Recerca i Turisme and the Conselleria d'Educaci\'o i Universitat del Govern de les Illes Balears,
the Conselleria d'Educaci\'o, Investigaci\'o, Cultura i Esport de la Generalitat Valenciana,
the National Science Centre of Poland,
the Swiss National Science Foundation (SNSF),
the Russian Foundation for Basic Research, 
the Russian Science Foundation,
the European Commission,
the European Regional Development Funds (ERDF),
the Royal Society, 
the Scottish Funding Council, 
the Scottish Universities Physics Alliance, 
the Hungarian Scientific Research Fund (OTKA),
the Lyon Institute of Origins (LIO),
the Paris \^{I}le-de-France Region, 
the National Research, Development and Innovation Office Hungary (NKFIH), 
the National Research Foundation of Korea,
Industry Canada and the Province of Ontario through the Ministry of Economic Development and Innovation, 
the Natural Science and Engineering Research Council Canada,
the Canadian Institute for Advanced Research,
the Brazilian Ministry of Science, Technology, Innovations, and Communications,
the International Center for Theoretical Physics South American Institute for Fundamental Research (ICTP-SAIFR), 
the Research Grants Council of Hong Kong,
the National Natural Science Foundation of China (NSFC),
the Leverhulme Trust, 
the Research Corporation, 
the Ministry of Science and Technology (MOST), Taiwan
and
the Kavli Foundation.
The authors gratefully acknowledge the support of the NSF, STFC, INFN and CNRS for provision of computational resources.
Research by DJS is supported by NSF grants AST-1821987, AST-1821967, AST-1813708, and AST-1813466.
We thank the Las Cumbres Observatory (LCO) and its staff for its continuing support of the ASAS-SN project.   ASAS-SN is supported by the Gordon and Betty Moore Foundation through grant GBMF5490 to the Ohio State University and NSF grant AST-1515927. Development of ASAS-SN has been supported by NSF grant AST-0908816, the Mt. Cuba Astronomical Foundation, the Center for Cosmology and AstroParticle Physics at the Ohio State University, the Chinese Academy of Sciences South America Center for Astronomy (CASSACA), the Villum Foundation, and George Skestos. KZS and CSK are supported by NSF grants AST-1515876, AST-1515927, and AST-1814440. Support for JLP is provided in part by FONDECYT through the grant 1191038 and by the Ministry of Economy, Development, and Tourism's Millennium Science Initiative through grant IC120009, awarded to The Millennium Institute of Astrophysics, MAS. 
Research  by SV  is  supported  by  NSF  grants  AST-1813176.
This document has been assigned LIGO Laboratory document number \ligodoc.

\end{acknowledgments}

\bibliography{paper.bib}

  \let\author\myauthor
  \let\affiliation\myaffiliation
  \let\maketitle\mymaketitle
  \title{Authors}
  \author{B.~P.~Abbott}
\affiliation{LIGO, California Institute of Technology, Pasadena, CA 91125, USA}
\author{R.~Abbott}
\affiliation{LIGO, California Institute of Technology, Pasadena, CA 91125, USA}
\author{T.~D.~Abbott}
\affiliation{Louisiana State University, Baton Rouge, LA 70803, USA}
\author{S.~Abraham}
\affiliation{Inter-University Centre for Astronomy and Astrophysics, Pune 411007, India}
\author{F.~Acernese}
\affiliation{Dipartimento di Farmacia, Universit\`a di Salerno, I-84084 Fisciano, Salerno, Italy}
\affiliation{INFN, Sezione di Napoli, Complesso Universitario di Monte S.Angelo, I-80126 Napoli, Italy}
\author{K.~Ackley}
\affiliation{OzGrav, School of Physics \& Astronomy, Monash University, Clayton 3800, Victoria, Australia}
\author{C.~Adams}
\affiliation{LIGO Livingston Observatory, Livingston, LA 70754, USA}
\author{V.~B.~Adya}
\affiliation{OzGrav, Australian National University, Canberra, Australian Capital Territory 0200, Australia}
\author{C.~Affeldt}
\affiliation{Max Planck Institute for Gravitational Physics (Albert Einstein Institute), D-30167 Hannover, Germany}
\affiliation{Leibniz Universit\"at Hannover, D-30167 Hannover, Germany}
\author{M.~Agathos}
\affiliation{Theoretisch-Physikalisches Institut, Friedrich-Schiller-Universit\"at Jena, D-07743 Jena, Germany}
\affiliation{University of Cambridge, Cambridge CB2 1TN, United Kingdom}
% ...
\collaboration{The LIGO Scientific Collaboration and the Virgo Collaboration}

\author{T.~W.-S.~Holoien}
\affiliation{Carnegie Observatories, 813 Santa Barbara Street, Pasadena, CA 91101, USA}
\author{C.~S.~Kochanek}
\affiliation{Department of Astronomy, The Ohio State University, 140 West 18th Avenue, Columbus, OH 43210, USA}
\affiliation{Center for Cosmology and AstroParticle Physics, The Ohio State University, 191 W. Woodruff Ave., Columbus, OH 43210, USA}
\author{J.~L.~Prieto}
\affiliation{N\'ucleo de Astronom\'ia de la Facultad de Ingenier\'ia, Universidad Diego Portales, Av. Ej\'ercito 441, Santiago, Chile}
\affiliation{Millennium Institute of Astrophysics, Santiago, Chile}
\author{B.~J.~Shappee}
\affiliation{Institute for Astronomy, University of Hawai'i, 2680 Woodlawn Drive, Honolulu, HI 96822, USA}
\author{K.~Z.~Stanek}
\affiliation{Department of Astronomy, The Ohio State University, 140 West 18th Avenue, Columbus, OH 43210, USA}
\affiliation{Center for Cosmology and AstroParticle Physics, The Ohio State University, 191 W. Woodruff Ave., Columbus, OH 43210, USA}
\collaboration{The ASAS-SN Collaboration}

\author{J.~Haislip}
\affiliation{Department of Physics and Astronomy, University of North Carolina at Chapel Hill, Chapel Hill, NC 27599, USA}
\author{V.~Kouprianov}
\affiliation{Department of Physics and Astronomy, University of North Carolina at Chapel Hill, Chapel Hill, NC 27599, USA}
\affiliation{Central (Pulkovo) Observatory of Russian Academy of Sciences, 196140 Pulkovskoye Ave. 65/1, Saint Petersburg, Russia}
\author{D.~E.~Reichart}
\affiliation{Department of Physics and Astronomy, University of North Carolina at Chapel Hill, Chapel Hill, NC 27599, USA}
\author{D.~J.~Sand}
\affiliation{Department of Astronomy/Steward Observatory, 933 North Cherry Avenue, Room N204, Tucson, AZ 85721-0065, USA}
\author{L.~Tartaglia}
\affiliation{Department of Astronomy and The Oskar Klein Centre, AlbaNova University Center, Stockholm University, SE-106 91 Stockholm, Sweden}
\author{S.~Valenti}
\affiliation{Department of Physics, University of California, 1 Shields Avenue, Davis, CA 95616-5270, USA}
\author{S.~Wyatt}
\affiliation{Department of Astronomy/Steward Observatory, 933 North Cherry Avenue, Room N204, Tucson, AZ 85721-0065, USA}
\author{S.~Yang}
\affiliation{Department of Physics, University of California, 1 Shields Avenue, Davis, CA 95616-5270, USA}
\affiliation{Department of Physics and Astronomy Galileo Galilei, University of Padova, Vicolo dell'Osservatorio, 3, I-35122 Padova, Italy}
\affiliation{INAF Osservatorio Astronomico di Padova, Vicolo dell’Osservatorio 5, I-35122 Padova, Italy}
\collaboration{The DLT40 Collaboration}

\author{F.~Salemi}
\affiliation{Albert-Einstein-Institut, Max-Planck-Institut f\"ur Gravitationsphysik, D-30167 Hannover, Germany}

  \newpage
  \maketitle

\end{document}